\documentclass[a4paper,11pt]{article}
\pdfoutput=1 

\usepackage{jheppub} 

\usepackage[T1]{fontenc} 
\usepackage{subcaption}
\usepackage{verbatim}
\usepackage{mathrsfs}
\usepackage{physics}
\usepackage{amsthm}
\usepackage{mathtools}
\usepackage{enumerate}

\newcommand{\Del}{\nabla}
\newcommand{\del}{\partial}

\renewcommand{\tilde}{\widetilde}
\renewcommand{\bar}{\overline}

\newcommand{\reals}{\mathbb{R}}
\newcommand{\comps}{\mathbb{C}}

\newcommand{\A}{\mathcal{A}}
\newcommand{\M}{\mathcal{M}}
\renewcommand{\O}{\mathcal{O}}
\renewcommand{\H}{\mathcal{H}}

\newcommand{\B}{\mathcal{B}}
\newcommand{\D}{\mathcal{D}}
\newcommand{\R}{\mathcal{R}}

\renewcommand{\L}{\mathcal{L}}

\newtheorem{theorem}{Theorem}

\theoremstyle{definition}
\newtheorem{definition}[theorem]{Definition}
\newtheorem{remark}[theorem]{Remark}

\renewcommand{\hat}{\widehat}

\numberwithin{theorem}{section}

\title{Bootstrap 2024: Lectures on ``The algebraic approach: when, how, and why?''}
\author{Jonathan Sorce}
\abstract{These lecture notes formed the basis of a mini-course on algebraic quantum field theory presented at the Bootstrap 2024 conference in Madrid.
The goal of the notes is to explain the merits of algebraic quantum field theory to a broad audience, and to explore (i) how to know when a particular question calls for an algebraic answer, (ii) how to apply the tools of algebraic QFT to such a problem, and (iii) why quantum field theorists of all stripes stand to benefit from a basic knowledge of algebraic quantum fields.
The first lecture focuses on the big-picture motivation behind the algebraic approach to quantum field theory, and an elaboration of its basic tools.
Subsequent lectures explain some of the modern achievements of the algebraic toolkit, including an argument for the averaged null energy condition and an enhanced understanding of entropy for semiclassical black holes.}

\makeatletter
\gdef\@fpheader{\vspace{2em}}
\makeatother

\begin{document}
\maketitle

\section{On the merits of algebraic quantum field theory}

Algebraic quantum field theory began in the late 1950s/early 1960s as an offshoot of axiomatic quantum field theory.
Both disciplines lie on the border of theoretical physics and pure mathematics, and both are concerned with the problem of formulating quantum field theory in a mathematically precise way.
Because of this history, algebraic quantum field theory is often viewed from the outside as a branch of mathematics, a domain of rigorous constructions that might be of conceptual interest, but that are of little practical consequence.

Those of us viewing it from the inside, however, see things quite differently.
The chief merit of algebraic quantum field theory is not that it provides a framework for proving theorems, but that it provides a way of \textit{thinking} about quantum field theory that lends itself to novel insights.
Viewing quantum field theory in algebraic terms can lead you to ask questions that you had not previously considered, and can lead you to creative solutions to existing problems that seemed insurmountable from other points of view.

The goal of these lecture notes is to present an audience of bootstrappers with the fundamental tools that they will need to think about quantum field theory algebraically.
My aim is not to suggest that algebraic frameworks are superior to other frameworks for quantum field theory; rather, I believe that having a plurality of ways of thinking about a single problem makes it easier and more exciting to make progress.
To that end, I emphasize that these lectures are not about the rigorous axioms of algebraic quantum field theory, but rather about what is often called the algebraic \textit{approach}.
This is an approach to problems in quantum field theory that shines in some settings where other approaches get stuck, and that gets stuck in some settings where other approaches shine.

Before we begin, it will be helpful to outline the algebraic approach by comparing it to the bootstrap philosophy.
The bootstrap philosophy takes conformal field theories as its chief objects of interest, both because they appear explicitly in statistical physics and in string theory, and because it is often possible to think of a non-conformal quantum field theory as a relevant deformation of a conformal one.
Conformal symmetry is taken as an organizing principle, and generic local operators are expressed in terms of ``primaries'' and ``descendants'' that transform irreducibly under the action of the conformal group.
The state-operator correspondence furnishes a bijection between these preferred local operators and the energy eigenstates of the Hamiltonian.
Questions of physical consequence are translated into questions about correlation functions of primary/descendant operators, and conformal invariance is used to show that all such correlation functions are completely determined by the spectrum of primaries and by the Euclidean three-point functions of primary operators (the OPE coefficients).
Once it is clear that every physical question about a CFT can be translated into a question about its spectrum and its OPE coefficients, the bootstrap philosophy focuses its effort on these quantities and how they are constrained by general principles such as crossing symmetry.

A key feature of the conformal bootstrap, and one that differs from textbook approaches to quantum field theory, is that it takes place in position space.
Position-space correlations, not S-matrix amplitudes, are the key objects of interest.
This perspective is also at the core of the algebraic approach.
The algebraic approach asks questions like, ``What are the observables that can be measured in a particular region of spacetime?'' and, ``In a given state, what are the general features of those observables?''
At a technical level, these questions are framed as questions about the \textit{algebras of fields localized to certain spacetime regions}.
The set of all fields localized in a spacetime region $\O$ generates an algebra $\A(\O),$ and questions about the structure of this algebra --- and the expectation values of its operators in states of physical interest --- form the core of the algebraic approach.

One key departure from the bootstrap philosophy is that the algebraic approach does not use any input from conformal symmetry (though local Lorentz invariance is often assumed), and does not in fact rely in any way on the structure of Minkowski spacetime.
While this generality has the disadvantage of discarding many of the powerful tools of conformal field theory, it has the advantage of allowing us to see general structural features of quantum field theory that do not rely on the specifics of Minkowski spacetime or of conformal symmetry.
This makes the algebraic approach particularly well suited for asking questions about cosmology and black holes.

The \hyperref[sec:lecture-1]{first lecture} explains the basics of the algebraic approach, including $*$-algebras and von Neumann algebras of fields.
It develops the theory of modular flow and explains the Unruh effect in an algebraic language.

The \hyperref[sec:lecture-2]{second lecture} introduces the theory of relative entropy in quantum field theory, and outlines the algebraic argument for the averaged null energy condition given in \cite{Faulkner:ANEC}.

The \hyperref[sec:lecture-3]{third lecture} describes some recent developments in using the algebraic approach to understand the entropy of semiclassical black holes.

Much more information about information-theoretic considerations in the algebraic approach can be found in the review articles \cite{Witten:entanglement} and \cite{Sorce:notes}.
A complete account of the mathematical foundations of the algebraic approach is provided in the textbook \cite{Haag:local}.
A modern perspective on the difference between an abstract algebra of fields and a particular Hilbert space representation can be found in \cite{Hollands-review}.

\section{Lecture 1: The fundamentals of the algebraic approach}
\label{sec:lecture-1}

\subsection{Preliminaries: correlation functions and operators}

What is a quantum field?
Probably the most conservative point of view is that a quantum field $\phi(x)$ is an object that can be placed in a correlation function to define quantities like
\begin{equation}
	\langle \Omega | \phi(x_1) \dots \phi(x_n) | \Omega\rangle.
\end{equation}
In Euclidean space, these correlation functions tend to actually be \textit{functions} that are defined away from coincident points; for example, the Euclidean two-point function of a primary field of dimension $\Delta$ in the vacuum state is
\begin{equation}
	\langle \Omega | \phi(x) \phi(y) |\Omega \rangle
		= \frac{C}{|x-y|^{2 \Delta}}.
\end{equation}
When this expression is analytically continued to Lorentzian signature, singularities appear not just at coincident points $x=y$, but also when $x$ and $y$ are null separated.
Because these null singularities are resolved when a small Euclidean time direction is added to the metric, the Minkowski-spacetime correlation functions are typically interpreted as boundary values of the analytically continued Euclidean correlators, and we write expressions like
\begin{equation}
	\langle \Omega | \phi(x_0, \vec{x}) \phi(y_0, \vec{y}) |\Omega \rangle
	= \frac{C}{\left( - (x_0 - y_0 - i \epsilon)^2 + (\vec{x} - \vec{y})^2 \right)^{2 \Delta}}.
\end{equation}
At a technical level, this is a ``distribution'' --- an expression that makes sense when we integrate against compactly supported smooth functions of $x$ and $y$, and defined by
\begin{align} \label{eq:iepsilon}
	\begin{split}
	& \int d^{d+1}x\, d^{d+1}y\, f(x) g(y) \langle \Omega | \phi(x) \phi(y) |\Omega \rangle \\
		& \qquad \qquad \equiv \lim_{\epsilon \to 0^+} \int dx_0\, dy_0\, d\vec{x}\, d\vec{y}\, f(x_0, \vec{x}) g(y_0, \vec{y}) \frac{C}{\left( - (x_0 - y_0 - i \epsilon)^2 + (\vec{x} - \vec{y})^2 \right)^{2 \Delta}}.
	\end{split}
\end{align}

The above point of view --- that a quantum field theory is a recipe for computing smeared correlation functions --- is perfectly sufficient for answering many questions.
The interesting quantities are correlations of fields, and because any measuring device has finite spatial resolution, the only quantities that are ever actually measured in quantum field theories are smeared correlation functions like the ones in equation \eqref{eq:iepsilon}.

Focusing only on correlation functions can be limiting, however, as this perspective obscures the \textit{quantum} nature of quantum field theory.
So far the state $|\Omega\rangle$ has been considered as some abstract object in which correlations can be defined, and not as a proper state in a vector space that can be superposed with other states.
Likewise, the fields $\phi(x)$ have been considered as abstract objects that furnish correlations, but not as operators that can change states, or for which a spectral decomposition can be performed.
To get at the quantum nature of quantum field theory, we would really like for $|\Omega\rangle$ to be a vector in some Hilbert space, and we would like for $\phi(x)$ to be an operator acting on that Hilbert space, such that we have the identity
\begin{equation}
	\langle \Omega | \phi(x) \phi(y) |\Omega \rangle = \langle \phi^{\dagger}(x) \Omega| \phi(y) \Omega \rangle.
\end{equation}

In the form stated above, this is not possible.
For example, if $\phi$ is a \textit{real} primary field, then for spacelike separated points $x$ and $y$, we are supposed to have
\begin{equation}
	\langle \phi(x) \Omega| \phi(y) \Omega \rangle
		= \frac{C}{|x-y|^{2 \Delta}}.
\end{equation}
Because this diverges in the limit $y \to x,$ the ``state'' $\phi(x) |\Omega\rangle$ is not normalizable, and cannot really be considered as a vector in Hilbert space.
To resolve this singularity, we must instead consider the ``smeared field''
\begin{equation}
	\phi[f]
		\equiv \int d^{d+1} x f(x) \phi(x).
\end{equation}
Interestingly, in Euclidean signature, the quantity
\begin{equation}
	\langle \phi[f] \Omega | \phi[f] \Omega \rangle
		\equiv \int d^{d+1} x\, d^{d+1}y\, f(x) f(y) \langle \Omega|\phi(x) \phi(y) |\Omega \rangle
\end{equation}
can be finite or infinite depending on the dimension $\Delta$, making the operator interpretation of the field confusing.
But in Lorentzian signature, the expression \eqref{eq:iepsilon} is always finite, and there is no issue with interpreting $\phi[f]$ as an operator.\footnote{See e.g. \cite[section 2]{Witten:smearing} for a pedagogical explanation of this finiteness.}

Generally it is fine to think of the difference between bare field operators $\phi(x)$ and smeared field operators $\phi[f]$ as a technical one.
The point is that in any good quantum field theory, there should exist observables localized to arbitrarily small regions of (Lorentzian) spacetime, and a Hilbert space on which they act.
In fact, given a self-consistent family of (Lorentzian) correlation functions $\langle \Omega | \phi(x_1) \dots \phi(x_n) |\Omega \rangle,$ one can always \textit{construct} a Hilbert space $\H$ with field operators $\phi[f]$ for which the expectation values reproduce the specified correlation functions.
The basic idea is to define an abstract vector space $V$ as the span of all objects of the form
\begin{equation}
	\phi[f_1] \dots \phi[f_n] |\Omega\rangle,
\end{equation}
to endow it with the inner product
\begin{equation}
	\langle \phi[g_1] \dots \phi[g_n] \Omega | \phi[f_1] \dots \phi[f_n] \Omega \rangle_V
		\equiv \langle \Omega | \phi^{\dagger}[g_1] \dots \phi^{\dagger}[g_{m}] \phi[f_1] \dots \phi[f_n] |\Omega\rangle_{\text{correlation function}},
\end{equation}
and to define $\H$ as the completion of $V$ with respect to this inner product, possibly after taking the quotient by a subspace of null states.
This procedure is called the ``GNS construction,'' or, in quantum field theory, the ``Wightman reconstruction theorem.''
For more detail on this construction, see e.g. \cite[chapter 7]{conway2000course}; for our purposes, it is enough to know that the ``correlation function'' perspective on quantum field theory and the ``field operator'' perspective on quantum field theory are equivalent.

\subsection{$*$-algebras and von Neumann algebras}

What information about a quantum field theory is accessible in a particular region of spacetime?
In the algebraic perspective, this is easy to answer: in the open subset $\O$ of Minkowski spacetime, we can access arbitrary observables generated by fields $\phi(x)$ with $x \in \O.$
Because we are asking about \textit{observable} fields, if there are any fermionic fields in our theory, then the observables associated with $\O$ should be constructed from composite bosonic combinations of the fermionic fields.
So we introduce the symbol $\A_0(\O)$ to denote the algebra of operators generated by arbitrary smearings $\phi[f]$ of the theory's bosonic fields, with $f$ a function supported in the spacetime region $\O$.\footnote{We typically have in mind that this includes \textit{all} local bosonic fields, not just some putative set of ``fundamental fields'' --- so in a conformal field theory, this would include primary fields, descendants, and appropriately renormalized composite objects constructed from those basic ingredients.}
\begin{definition}
	Given a spacetime region $\O$, $\A_0(\O)$ is the set of all polynomial combinations of smeared bosonic fields with support in $\O$.
\end{definition}

By this definition, the set $\A_0(\O)$ is an \textit{algebra}, i.e., a set of operators that is closed under multiplication, addition, and scalar multiplication.
In fact, it should be considered as a $*$-algebra, which is an algebra with an adjoint operation.
This is because for any field $\phi(x)$ there should also be an adjoint field $\phi^{\dagger}(x),$ and the algebra $\A_0(\O)$ contains arbitrary operators generated from all combinations of these fields.

\begin{remark}
	$\A_0(\O)$ is a $*$-algebra, with adjoint operation
	\begin{equation}
		\phi^{\dagger}[f] = \left( \int d^{d+1}x\, \phi(x) f(x) \right)^{\dagger} = \int d^{d+1}(x) \phi^{\dagger}(x) \bar{f}(x).
	\end{equation}
\end{remark}

\begin{figure}
	\centering
	\includegraphics[scale=1.5]{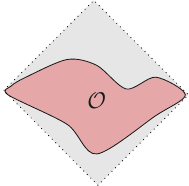}
	\caption{If a quantum field theory is causally well behaved, then we expect that operators localized to a region $\O$ can be used to construct any operator in the domain of dependence of $\O$.}
	\label{fig:O-domain}
\end{figure}

We are now faced with an interesting conceptual question: does $\A_0(\O)$ contain all operators that can be measured in the spacetime region $\O$?
The answer to this is almost certainly no.
For example, due to Lorentz invariance, we should be able to construct within $\O$ any field operator localized within the domain of dependence of $\O$; see figure \ref{fig:O-domain}.
But this is not possible strictly within the algebra $\A_0(\O).$
Instead we should really consider some completed algebra $\A(\O)$, which consists of all limits of operators in $\A_0(\O)$ in an appropriate topology.
In fact, it is most convenient to construct $\A(\O)$ as an algebra of \textit{bounded} operators, whereas the operators in $\A_0(\O)$ are typically unbounded --- so to get $\A(\O)$ from $\A_0(\O)$ we will need to come up with some two-step process where we first take appropriate bounded functions of operators in $\A_0(\O)$, then complete the resulting space of operators in a topology of our choosing.

We will sidestep the thorny details of such a construction by saying that $\A(\O)$ should contain at least all those operators that commute with all field operators in the spacelike complement of $\O$.
I.e., if $\O'$ denotes all points spacelike separated from $\O$, then we should have\footnote{It is perfectly sensible to say that a bounded operator commutes with an unbounded one; this means that the commutator is defined and vanishes for all vectors in the domain of the unbounded operator.}
\begin{equation}
	\A(\O) \supseteq \{T \text{ bounded } | [T, a'] = 0 \text{ for } a' \in \A_0(\O')\}.
\end{equation}
As a minimal guess for what $\A(\O)$ should be, we will simply declare equality:\footnote{There are a few technical subtleties suppressed in this equation, namely that we will take $\A(\O)$ to consist only of bounded operators, even though the operators in $\A_0(\O')$ may be unbounded.
For an introduction to unbounded operators, see \cite[section 2]{Sorce:intuitive}.}
\begin{equation}
	\A(\O) \equiv \{T \text{ bounded } | [T, a'] = 0 \text{ for } a' \in \A_0(\O')\}.
\end{equation}
We will call this the \textbf{von Neumann algebra} of operators associated with $\O.$
We think of it as the set of all observables that can be accessed in the spacetime region $\O.$

A few remarks before we proceed, though none of them will be essential in what follows:
\begin{itemize}
	\item 
	It is a fairly easy exercise to show that $\A(\O)$ is closed in the strong topology, meaning that if there is a sequence of operators $T_n$ in $\A(\O)$ such that $T_n |\psi\rangle$ converges to $T |\psi\rangle$ for every state $|\psi\rangle$ in Hilbert space, then $T$ is in $\A(\O).$
	In spite of this, it is not necessarily true that $\A(\O)$ is the closure of (an appropriate bounded version of) $\A_0(\O)$ with respect to this topology.
	We generally have
	\begin{equation}
		\A_0(\O)_{\text{bounded}} \subseteq \bar{\A_0(\O)_{\text{bounded}}} \subseteq \A(O).
	\end{equation}
	Sometimes, in gauge theories, whether or not the second inclusion is strict for topologically nontrivial regions depends on whether or not you include Wilson loops among the fields generating $\A_0(\O).$
	For recent discussion on this point, see \cite{Casini:completeness}.
	\item While the construction of the algebra $\A(\O)$ depended on having a Hilbert space hanging around, the algebra $\A_0(\O)$ could have been considered as some abstract algebra of field symbols, without thinking of these symbols as operators acting on a Hilbert space.
	This can be a useful perspective when thinking about the fact that quantum field theories typically have multiple inequivalent Hilbert space representations --- for example, in Minkowski spacetime, the space of excitations above a thermal state and the space of excitations above the vacuum are unitarily inequivalent.
	Axiomatic approaches to quantum field theory in curved spacetime, such as \cite{Hollands:axiomatic}, generally take \textit{abstract} $*$-algebras of fields $\A_0(\O)$ as the starting point for defining the theory.
	They then determine families of physically acceptable correlation functions for these algebras, and introduce Hilbert spaces via the GNS construction alluded to at the end of the previous subsection.
\end{itemize}

\subsection{Information theory, thermodynamics, and modular flow}

When studying subsystems of a many-body system on a lattice in the absence of gauge constraints, the typical object of study is a density matrix.
Concretely, given a factorizing Hilbert space $\H = \H_A \otimes \H_B,$ and a state $|\Psi\rangle_{AB},$ we define the density matrix
\begin{equation}
	\rho_{A}
		= \tr_{B} |\Psi\rangle\langle\Psi|_{AB}.
\end{equation}
If the state $|\Psi\rangle$ is sufficiently entangled between $A$ and $B$, then it will be possible to take the logarithm of $\rho_A,$ and we can write
\begin{equation} \label{eq:Gibbs-rho}
	\rho_{A}
		= e^{- (- \log{\rho_A})}.
\end{equation}
Because the density matrix satisfies $0 < \rho_A \leq 1,$ the operator
\begin{equation}
	K_A
		\equiv - \log{\rho}_{A}
\end{equation}
is Hermitian and bounded below.
Consequently, it can be thought of at the formal level as a ``Hamiltonian,'' though perhaps not one of independent physical interest.
It is called the \textbf{modular Hamiltonian} of $\rho_{A}.$
Equation \eqref{eq:Gibbs-rho} expresses that $\rho_{A}$, which started its life as a generic full-rank density matrix, is a Gibbs state with respect to the modular Hamiltonian.

This way of thinking can be quite useful.
In statistical mechanics, typically we pick a Hamiltonian of interest, such as one describing nearest-neighbor interactions on the lattice, and study the Hamiltonian by studying the statistical properties of its Gibbs states.
Quantum information theory goes the other way around --- it asks questions about a generic full-rank state by thinking of it as a Gibbs state for its modular Hamiltonian, and applying the principles of statistical mechanics.
This is the origin of von Neumann's famous entropy formula
\begin{equation}
	S(\rho_A) \equiv - \tr\left( \rho_A \log \rho_A\right).
\end{equation}

\begin{figure}
	\centering
	\includegraphics[scale=1]{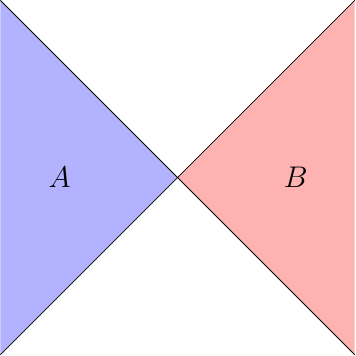}
	\caption{A pair of complementary spacetime regions $A$ and $B$, containing commuting algebras $\A(A)$ and $\A(B)$.
	In the presence of a regulator, we may think of the Hilbert space as factorizing in the form $\H = \H_A \otimes \H_B.$}
	\label{fig:complementary-spacetime}
\end{figure}

It would be nice to apply these ideas to local algebras in quantum field theory to understand the correlations present in physical states.
Unfortunately, in quantum field theory, there are no density matrices associated with subregions of spacetime.
Ultraviolet divergences coming from short-distance entanglement render every density matrix non-normalizable.
For some problems this issue can be dealt with by working on a lattice and taking the limit as the cutoff is removed, but for other calculations the ultraviolet issues are so severe that it is more profitable to try an alternate route.
The key idea --- see figure \ref{fig:complementary-spacetime} --- is that if $A$ and $B$ are complementary spacetime regions, and $|\Psi\rangle_{AB}$ is a global state, then while the density matrices $\rho_A$ and $\rho_B$ might be individually UV-divergent, the combination
\begin{equation} \label{eq:regulated-Delta}
	\Delta_{\Psi} \equiv \rho_{A} \otimes \rho_{B}^{-1} 
\end{equation} 
may be well defined in the continuum, as might its logarithm
\begin{equation} \label{eq:regulated-K}
	K_{\Psi} \equiv - \log{\Delta_{\Psi}} = K_A - K_B.
\end{equation}
The operator $\Delta_{\Psi}$ --- if it exists --- is called the \textbf{modular operator} for the region $A$ in the state $|\Psi\rangle_{AB}$.
The operator $K$ is called the \textbf{full modular Hamiltonian}, though often the word ``full'' is dropped and it is simply called the modular Hamiltonian.

A powerful mathematical technology called \textit{Tomita-Takesaki theory} \cite{takesaki2006tomita, takesaki-vol2} allows modular operators and full modular Hamiltonians to be defined for a broad class of states and subregions in continuum quantum field theory.
These operators are completely well defined in the continuum theory without any regulator, and reproduce the formulas \eqref{eq:regulated-Delta} and \eqref{eq:regulated-K} when a regulator is introduced.
We will explore a little bit of this theory below; a complete account of the basic theory for physicists can be found in \cite{Sorce:intuitive}, and many interesting applications are discussed in \cite{Witten:entanglement}.

In order for both equations \eqref{eq:regulated-Delta} and \eqref{eq:regulated-K} to make sense, a state $|\Psi\rangle_{AB}$ must have the property that its reduced density matrices $\rho_{A}$ and $\rho_{B}$ are full rank.
So the quantum field theory states for which a modular operator and modular Hamiltonian make sense should be those that have ``full rank density matrices.''
Of course, since density matrices are UV-divergent in quantum field theory, we need a precise way of capturing the notion of a ``full rank density matrix'' that uses only the mathematical structure that is actually present in QFT --- namely, the von Neumann algebras $\A(\O)$ associated with spacetime regions $\O.$

\begin{definition}
	A QFT state $|\Psi\rangle$ is \textbf{cyclic} with respect to a von Neumann algebra $\A$ if the states
	\begin{equation}
		\{a |\Psi\rangle\, |\, a \in \A\}
	\end{equation}
	are dense in Hilbert space.
\end{definition}

\begin{remark}
	For a factorizing Hilbert space $\H = \H_A \otimes \H_B,$ it is easy to see that a state $|\Psi\rangle_{AB}$ has full-rank reduced density matrix on $\H_B$ if and only if it is cyclic with respect to the algebra of operators acting nontrivially on $\H_A$ and acting as the identity on $\H_B$.
	So ``cyclic'' states generalize those with full-rank reduced density matrices to settings where no density matrices exist.
\end{remark}

Based on the above definition and remark, we see that for a region $\O$ in spacetime, we expect to be able to define a modular Hamiltonian and modular operator for the state $|\Psi\rangle$ with respect to region $\O$ if and only if $|\Psi\rangle$ is cyclic with respect to both $\A(\O)$ and the complementary algebra $\A(\O').$
Such a state is called \textit{cyclic and separating} with respect to $\A(\O)$, based on the following definition and remark.

\begin{definition}
	A state $|\Psi\rangle$ is \textbf{separating} with respect to a von Neumann algebra $\A$ if the only operator in $\A$ that annihilates $|\Psi\rangle$ is the zero operator.
\end{definition}

\begin{remark}
	It is a fairly straightforward exercise to show that a state is cyclic with respect to a von Neumann algebra if and only if it is separating with respect to the von Neumann algebra of operators that commute with the elements of the original algebra.
\end{remark}

There are many cyclic and separating states in quantum field theory, as explained in the following remark.

\begin{remark}
	The \textit{Reeh-Schlieder theorem} \cite{reeh1961bemerkungen} shows, within the framework of the Wightman axioms for quantum field theory, that the vacuum state of Minkowski spacetime is cyclic for any open subset of spacetime.
	The proof uses certain analytic properties of vacuum correlation functions, and can easily be extended to arbitrary states of bounded energy-momentum; see e.g. \cite[section 2.3]{Witten:entanglement}.
	In more general spacetimes, every field theory is expected to have an abundance of states that are cyclic for any choice of region \cite{Sanders:reeh}.
	The construction of states that are cyclic for \textit{every} region simultaneously is mathematically difficult, but it seems plausible that they should exist, and that many states of physical interest are cyclic and separating for every spacetime region.
\end{remark}

Now that we have come up with a good definition of states for which we expect to be able to define modular operators and modular Hamiltonians, we simply quote Tomita's theorem, which says that these operators exist.

\begin{theorem}[Tomita's theorem]
	Given a von Neumann algebra $\A$ and a state $|\Psi\rangle$ that is cyclic and separating for $\A$, there exists an operator $\Delta_{\Psi},$ called the modular operator, with the following properties:
	\begin{itemize}
		\item $\Delta_{\Psi}$ is Hermitian with nonnegative spectrum and it has no kernel, so its logarithm $K_{\Psi} = - \log \Delta_{\Psi}$ exists.
		\item While $\Delta_{\Psi}$ may be unbounded, and therefore unable to act on every vector in Hilbert space, the operator $\Delta_{\Psi}^{1/2}$ is able to act on those vectors of the form $a |\Psi\rangle$ for $a \in \A,$ and for such vectors we have the identity
		\begin{equation}
			\langle \Delta_{\Psi}^{1/2} a \Psi| \Delta_{\Psi}^{1/2} b \Psi\rangle
				= \langle b^{\dagger} \Psi | a^{\dagger} \Psi \rangle.
		\end{equation}
		In the case of a factorizing Hilbert space, one can easily check that these are consistent with the matrix elements of equation \eqref{eq:regulated-Delta}.
		\item The unitary flow generated by $K_{\Psi},$ called \textbf{modular flow}, satisfies
		\begin{equation}
			e^{i K_{\Psi} t} a e^{-i K_{\Psi} t} \in \A \quad \text{for} \quad a \in \A.
		\end{equation}
	\end{itemize}
\end{theorem}

In quantum field theory, we can use the modular operator to understand things that we would usually try to understand in lattice systems using density matrices.
The general technique is to prove some general property of modular operators, to look for a special QFT state in which the modular operator takes an interesting form, and to apply the general theorem to the specific case.
We will explore some examples of this in lectures 2 and 3.
In what remains of this lecture, we will try to get a better feel for the modular operator by computing it for the Rindler subalgebra of Minkowski spacetime in the vacuum state; the modular flow will be seen to be a boost preserving the Rindler wedge, which gives a mathematically precise statement of the Unruh effect.

\subsection{The Unruh effect}

The Unruh effect, as formulated in \cite{Unruh:effect}, states that an observer following a uniformly accelerating trajectory in Minkowski spacetime will experience the vacuum as a thermal bath.
In the previous subsection, we explained that every state looks thermal with respect to the flow of time generated by its modular Hamiltonian.
So what we really want to show is that the modular Hamiltonian of the vacuum state, restricted to a Rindler wedge, generates a uniformly accelerating transformation of that wedge --- i.e., a boost.

This statement was proved by Bisognano and Wichmann in \cite{Bisognano:1, Bisognano:2}, and in fact their proof predates Unruh's paper \cite{Unruh:effect}.
Interestingly, the connection between the two works was not immediately realized; according to iNSPIRE, the first paper to cite both works was \cite{Sewell:1982zz}, which was published six years later.

We will now present the calculation of the vacuum-Rindler modular Hamiltonian that was performed in \cite{Unruh:integral}, though updated to a more modern language.
One ideologically interesting feature of this calculation is that it is non-rigorous, as it uses Euclidean path integrals, density matrices, and a cavalier approach to ultraviolet renormalization.
This reflects a general philosophy among practitioners of the algebraic approach --- we are happy to know that the modular Hamiltonian is rigorously defined, and if we have to relax our standards of rigor to arrive at a prediction for the form of the modular Hamiltonian for a given state, we do not view this as a tragedy.
The fact that we know the object we are computing is well defined in the continuum gives us faith in the final destination of our mathematically imperfect journey.
To fill in the gap, though, I will sketch the rigorous proof due to Bisognano and Wichmann at the end of this section.

\begin{figure}
	\centering
	\begin{subfigure}[b]{0.48\textwidth}
		\centering
		\includegraphics[width=\textwidth]{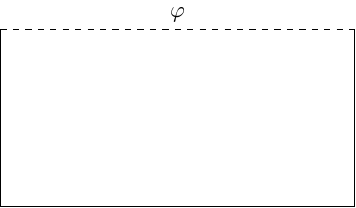}
		\caption{}
		\label{fig:vacuum-path-integral}
	\end{subfigure}
	\hfill
	\begin{subfigure}[b]{0.48\textwidth}
		\centering
		\includegraphics{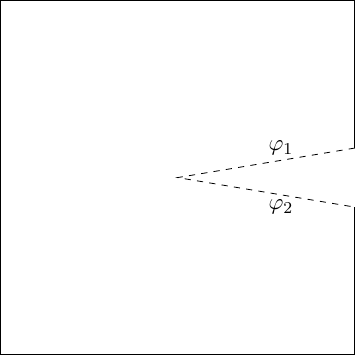}
		\caption{}
		\label{fig:vacuum-density-matrix}
	\end{subfigure}
	\caption{(a) The path integral on a Euclidean half-plane with boundary condition $\varphi$ is proportional to the overlap $\langle \varphi |\Omega\rangle.$
		(b) The path integral on a Euclidean plane with discontinuous boundary conditions $\phi_1$ and $\phi_2$ on $x \geq 0$ is proportional to $\langle\varphi_2|\rho_R|\varphi_1\rangle.$}
	\label{fig:vacuum-figures}
\end{figure}

Let $|\Omega\rangle$ be the vacuum state of Minkowski spacetime, and suppose we have chosen a regulation procedure such that we can tensor factorize the Hilbert space as $\H = \H_L \otimes \H_R,$ with $L$ the left Rindler wedge and $R$ the right Rindler wedge.
We might for example impose a ``hard wall'' cutoff on quantum fields near the surface where $L$ and $R$ meet.
In this regulated theory, $|\Omega\rangle$ has density matrices $\rho_L$ and $\rho_R$ on the two tensor factors, and the modular Hamiltonian associated to region $R$ is
\begin{equation} \label{eq:split-unruh-modham}
	K_{\Omega} = - \log{\rho_R} + \log{\rho_L}.
\end{equation}
Clearly to understand this operator it suffices to understand the density matrices $\rho_R$ and $\rho_L$.
The density matrix $\rho_R$ is written as
\begin{equation}
	\rho_R = \tr_L |\Omega\rangle\langle\Omega|.
\end{equation}
In the unregulated theory, the state $|\Omega\rangle$ can be represented up to a normalization constant as a Euclidean path integral over an infinite past, with open boundary conditions at a time slice; see figure \ref{fig:vacuum-path-integral}.
The density matrix $\rho_R$ is computed by taking two copies of this path integral and gluing them together along their boundaries for $x \leq 0$; see figure \ref{fig:vacuum-density-matrix}.
Concretely, this means that for a field configurations $\varphi_1$ and $\varphi_2$ on a slice of $R$, the matrix element $\langle \varphi_1 | \rho_R | \varphi_2 \rangle$ is computed by the Euclidean path integral shown in figure \ref{fig:vacuum-density-matrix}.

\begin{figure}
	\centering
	\begin{subfigure}[b]{0.48\textwidth}
		\centering
		\includegraphics[width=\textwidth]{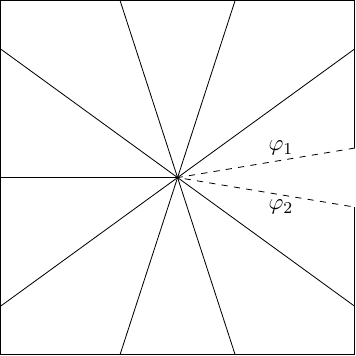}
		\caption{}
		\label{fig:unruh-path-integral}
	\end{subfigure}
	\hfill
	\begin{subfigure}[b]{0.48\textwidth}
		\centering
		\includegraphics{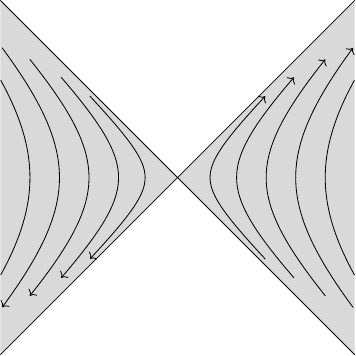}
		\caption{}
		\label{fig:boost}
	\end{subfigure}
	\caption{(a) An angular slicing of the path integral producing matrix elements of $\rho_R.$
		(b) A sketch of the boost that is future-directed in the right Rindler wedge.}
	\label{fig:unruh-figures}
\end{figure}

This path integral is generally divergent due to the possibility of a discontinuity between $\varphi_1$ and $\varphi_2$ on the $x=0$ surface sketched in figure \ref{fig:vacuum-density-matrix}.
We will nevertheless proceed by a formal manipulation of the path integral in figure \ref{fig:vacuum-density-matrix}, and trust that any apparent ultraviolet issues will sort themselves out by the time we are done.
The key insight is that this path integral can be ``re-sliced'' according to the angular slicing shown in figure \ref{fig:unruh-path-integral}, in which case it can be interpreted as the time-ordered exponential of the constant-angle generators\footnote{An explicit derivation of this formula is given in appendix \ref{app:angular-slicing}.}
\begin{equation} \label{eq:angle-generator}
	K_{\theta} = \int_{\text{constant $\theta$}} T^{E}_{\mu \nu} \left[ x \left( \frac{\del}{\del \tau} \right)^{\mu} + \tau \left(\frac{\del}{\del x}\right)^{\mu} \right] d \Sigma^{\nu},
\end{equation}
with $T^E$ the stress-energy tensor for the Euclidean action.
This implies an expression for $\rho_R$ of the form
\begin{equation}
	\rho_R \propto T \exp\left[ - \int_{0}^{2\pi} d\theta K_{\theta}\right].
\end{equation}
But so long as the field theory has a Euclidean rotation symmetry, the generator $K_{\theta}$ is conserved, so this simplifies to
\begin{equation}
	\rho_R \propto \exp\left[ - 2 \pi K_{\theta=0}\right]
		= \exp\left[- 2 \pi \int_{0}^{\infty} dx\, x\, T^{E}_{\tau \tau} \right].
\end{equation}
The Euclidean stress tensor $T^E$ and the Lorentzian stress tensor $T$ satisfy the equation $T^{E}_{\tau \tau} = T_{t t}$; so up to a (possibly infinite) normalization constant $Z$, we have
\begin{equation}
	\rho_R = \frac{1}{Z} \exp\left[- 2 \pi \int_{0}^{\infty} dx\, x\, T_{t t} \right].
\end{equation}
Repeating the same procedure for $\rho_L$ gives
\begin{equation}
	\rho_L = \frac{1}{Z} \exp\left[2 \pi \int_{-\infty}^{0} dx\, x\, T_{t t} \right],
\end{equation}
with the \textit{same} normalization constant due to the symmetry of the problem.
These formulas, plugged into equation \eqref{eq:split-unruh-modham}, give the modular Hamiltonian
\begin{equation} \label{eq:boost}
	K_{\Omega} = 2 \pi \int_{-\infty}^{\infty} dx\, x\, T_{tt}.
\end{equation}
This is exactly the operator that generates the boosts shown in figure \ref{fig:boost}, which are future-directed in the right Rindler wedge and past-directed in the left Rindler wedge.
Thus the Minkowski vacuum is thermal with respect to boosts, which is a precise statement of the Unruh effect.

While we will not discuss it any more in this mini-course, it is possible to prove rigorously that the modular Hamiltonian of the vacuum generates boosts, as was done by Bisognano and Wichmann in \cite{Bisognano:1, Bisognano:2}.
The idea is to use the observation that for any cyclic-separating state $|\Psi\rangle$ for a von Neumann algebra $\A$, the modular flow $e^{i K_{\Psi} s}$ satisfies a property known as the \textit{KMS condition}, which is a statement about the analytic structure of the correlation function
\begin{equation}
	F(s) = \langle \Psi | e^{i K_{\Psi} s} a e^{-i K_{\Psi} s} b |\Psi \rangle, \qquad a, b \in \A.
\end{equation}
We will not give an explicit statement of the KMS condition here, though see \cite{Sorce:intuitive} for a recent treatment; it is enough to know that the KMS condition is a statement about the analytic continuation of the above function in the $s$-plane.
A theorem due to Takesaki \cite[theorem 1.2]{takesaki-vol2} states that the KMS condition uniquely determines the modular Hamiltonian, as the modular Hamiltonian is the only Hermitian operator for which all unitarily-evolved correlation functions have KMS analytic structure.
Bisognano and Wichmann used the Wightman axioms to show that in the vacuum state, all boost-evolved two-point functions in a Rindler wedge have KMS analytic structure, which implies that the boost is the modular Hamiltonian for the vacuum state in the Rindler wedge.

\section{Lecture 2: Relative entropy and the ANEC}
\label{sec:lecture-2}

\subsection{History of the ANEC}

General relativity is the classical theory of Einstein's equation:
\begin{equation}
	R_{\mu \nu} - \frac{1}{2} R g_{\mu \nu} = 8 \pi T_{\mu \nu}. 
\end{equation}
This is a nonlinear PDE by which a spacetime metric $g_{\mu \nu}$ is determined from a source stress-energy tensor $T_{\mu \nu}.$
While a tensor $T_{\mu \nu}$ satisfying the above equation exists for \textit{any} metric $g_{\mu \nu}$, not every spacetime can be sourced by a stress-energy tensor that comes from physically reasonable matter.
Understanding the features of ``physically reasonable'' stress-energy tensors is therefore essential for understanding the geometric features of physically reasonable spacetimes.

There is a long history in general relativity of imposing very general \textit{energy conditions} on $T_{\mu \nu}$, and studying the consequences for solutions of Einstein's equations.
The simplest and most general of these is the \textit{null energy condition} (NEC), which requires that for any null vector $k^\mu$, We have
\begin{equation}
	T_{\mu \nu} k^\mu k^\nu \geq 0. \qquad \text{(NEC)}
\end{equation}
This energy condition is satisfied by nearly every classical field theory that one would like to consider; to be precise, in a classical field theory, $T_{\mu \nu}$ is a functional of field configurations, and for a large class of physically motivated theories, the NEC is satisfied for field configurations that solve the classical equations of motion.
Amazingly, the NEC is sufficient to prove many general constraints on physical spacetimes sourced by NEC-satisfying matter.
For example, the NEC is an important ingredient in the proofs of the singularity theorems \cite{Penrose:singularity, Hawking:singularity}, the area theorem \cite{Hawking:area}, topological censorship \cite{topological-censorship}, and the dynamical first law of black hole mechanics \cite{Gao:physical-first-law}.

There is, however, a problem.
While classical field theories typically satisfy the null energy condition, \textit{quantum} field theories do not.
Indeed, it can be shown that in any quantum field theory in Minkowski spacetime, for any point $x$ and any null vector $k^{\mu}$ at $x,$ there exist states in the vacuum sector for which the one-point function of $T_{\mu \nu} k^{\mu} k^{\nu}$ is negative at $x$ \cite{Epstein:negativity}.
This fact is essential to Hawking's derivation of black hole radiance \cite{Hawking:radiance} --- if quantum effects can cause a black hole to evaporate, then they must be able to violate the NEC, since black hole area cannot decrease if the NEC is satisfied.

What do we make of this?
Our universe has quantum fields, and these fields have a stress-energy, so they must make a contribution to our spacetime structure via Einstein's equations.
If these quantum fields can violate the NEC, do we have to throw away all of the general theorems that have been proved about spacetimes satisfying the NEC?
Do they say anything about our universe at all?

In trying to answer this question, several authors \cite{Tipler:ANEC-1, Roman:ANEC-2, Roman:ANEC-3, Borde:ANEC-4} discovered that for many applications to general relativity, the NEC is a stronger assumption than one actually needs.
Many theorems can be proved from the \textit{averaged null energy condition} (ANEC), which states that for any complete null geodesic $\gamma$ with tangent vector $k^a,$ we should have
\begin{equation}
	\int_{\gamma} T_{\mu \nu} k^{\mu} k^\nu \geq 0.
\end{equation}
This observation led relativists to start asking whether the ANEC might be satisfied at the quantum level in general curved spacetimes.
The answer to this question is no \cite{Visser:ANEC-violation, Urban:ANEC-violation}, as one can explicitly construct quantum field theory states in conformally flat spacetimes for which the one-point function of the ANEC operator is negative.\footnote{In fact, there is a simpler example in which the one-point function of the ANEC operator is negative, which is a two-dimensional cylinder in the vacuum state of a conformal field theory; this occurs due to the negative Casimir energy of the vacuum.
But this case is a bit pathological because a complete null geodesic on a two-dimensional cylinder is \textit{chronal}, meaning it can send timelike signals to itself; a slight generalization of the ANEC called the ``achronal averaged null energy condition'' (AANEC) rules out this counterexample, but not the ones from \cite{Visser:ANEC-violation, Urban:ANEC-violation}.}
Several authors \cite{Flanagan:self-consistent, Graham:self-consistent} have speculated about a further refinement, the \textit{self-consistent achronal averaged null energy condition} (SCAANEC), which has no known counterexamples and suffices to prove many of the fundamental theorems about general relativity.
The SCAANEC remains an area of active research.

Even if the ANEC is not true at the quantum level in general curved spacetimes, one can ask if it holds in certain special spacetimes, or in certain special states.
At the simplest level, one might ask: does the ANEC hold for quantum fields in Minkowski spacetime?
The answer to this question appears to be yes.
While a rigorous proof has not been produced,\footnote{\label{foot:rigor}Promising steps have been made toward a rigorous proof of the ANEC in \cite{Casini:ANEC, Faulkner:QNEC-1,Faulkner:QNEC-2} via an observed connection with older work on so-called half-sided modular inclusions \cite{Borchers:CPT, Wiesbrock:HSMI}.
Basically, one can show that for any positive smearing function on a Rindler horizon, there is a positive operator whose commutator with boosts is of the same algebraic form as the commutator of the corresponding smeared ANEC operator with boosts.
It seems to me however that more must be done to establish that this operator \textit{is} the smeared ANEC operator in theories where a smeared ANEC operator already exists, and not some distinct operator that happens to satisfy a similar algebraic relation.
The work of \cite{Faulkner:ANEC}, which is the subject of the rest of this lecture, provides a compelling argument for this identification using the path integral.} compelling arguments for the ANEC in Minkowski spacetime have been produced in two separate arguments using ideas from the bootstrap \cite{Hartman:ANEC} and ideas from the algebraic approach \cite{Faulkner:ANEC}.
These general arguments follow a long history of arguments for the ANEC in special cases \cite{Yurtsever:ANEC-proof, Klinkhammer:ANEC-proof,Wald:ANEC-proof}.

The original motivation for studying the ANEC came from general relativity; from this perspective, the question of whether the ANEC holds in Minkowski spacetime seems to be largely academic.
A significant departure from this perspective was made in \cite{Hofman:2008ar}, in which Hofman and Maldacena showed that the positivity of the ANEC operator in certain states can be used to prove general constraints on the structure of conformal field theories.
This philosophy has been used in e.g. \cite{Ceyhan:QNEC, Hartman:RG-1, Hartman:RG-2} to show that many interesting structural statements about quantum field theory are implied by positivity of the ANEC operator in Minkowski spacetime.
Generalizations of the ANEC operator known as light-ray operators have been used for similar purposes \cite{Kravchuk:light-ray, Kologlu:light-ray, Korchemsky:light-ray}.

In the rest of this lecture, we will explore the algebraic argument for the ANEC given in \cite{Faulkner:ANEC}.
The actual argument involves many technical tricks and heuristic manipulations of formal expressions, so I do not think it is pedagogically useful to explain the paper in detail.
We will content ourselves with understanding the big-picture reasoning that motivates the calculation of \cite{Faulkner:ANEC}, and what it has to teach us about how we think about the ANEC.

\subsection{The ANEC as a transformation, and relative entropy}

\begin{figure}
	\centering
	\includegraphics{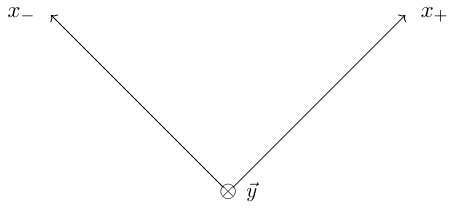}
	\caption{Light-cone coordinates $x_+$ and $x_-$ together with transverse, ``internal'' directions labeled by $\vec{y}$.}
	\label{fig:light-cone-coordinates}
\end{figure}

In $(d+1)$-dimensional Minkowski spacetime we can choose light-cone coordinates $x_{+}$ and $x_{-}$, together with transverse coordinates $\vec{y}$ --- see figure \ref{fig:light-cone-coordinates}.
The ``ANEC operator'' is the quantity
\begin{equation}
	\text{ANEC}(\vec{y})
		= \int_{-\infty}^{\infty} dx_+\, T_{++}(x_+, x_- = 0, \vec{y}). 
\end{equation}
We would like to ask if this quantity has positive expectation value in every quantum field theory state in Minkowski spacetime.

One useful way of thinking about the ANEC operator is in terms of the spacetime transformation that it generates.
Using the algebra of the stress tensor with matter fields, it is easy to see that for fields on the null ray where ANEC$(\vec{y})$ is supported, the operator ANEC$(\vec{y})$ generates a translation in the $x_+$ direction.
I.e., we have an equation like
\begin{equation}
	[\text{ANEC}(\vec{y}_1), \phi(x_+, x_-=0, \vec{y}_2)]
		\sim \delta(\vec{y}_1 - \vec{y}_2) \del_{x_+} \phi(x_+, x_-=0, \vec{y}_2).
\end{equation}
So what we are really asking is if this geometric transformation of fields is positively generated.

\begin{figure}
	\centering
	\includegraphics[scale=1.5]{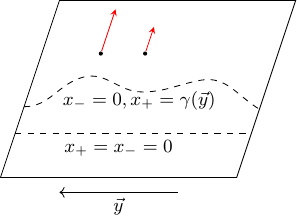}
	\caption{A view of a Rindler horizon, $x_-=0$, together with the curves $x_+ =0 $ and $x_+=\gamma(\vec{y})$.
	The black dots represent matter operators; the commutator of a matter operator with ANEC$[\gamma]$ generates a push-forward of the matter operator along its null generator, with the infinitesimal magnitude of the push-forward given by the magnitude of the function $\gamma$; cf. equation \eqref{eq:smeared-push}.}
	\label{fig:null-push}
\end{figure}

\begin{figure}
	\centering
	\includegraphics{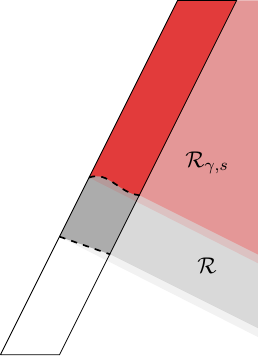}
	\caption{A Rindler wedge $\R$ and the sub-wedge $R_{\gamma,s}$ obtained by integrating the vector field $\gamma(\vec{y})\frac{\del}{\del x_+}$ by parameter $s$.}
	\label{fig:null-deformed-rindler-R}
\end{figure}

It is a bit easier to handle all of these expressions if we let $\gamma(\vec{y})$ be a positive function of $\vec{y}$, and define the smeared ANEC operator ANEC$[\gamma]$ by
\begin{equation}
	\text{ANEC}[\gamma]
	= \int_{-\infty}^{\infty} dx_+\, d\vec{y}\, \gamma(\vec{y})\, T_{++}(x_+, x_- = 0, \vec{y}). 
\end{equation}
Clearly ANEC$(\vec{y})$ is positive if and only if ANEC$[\gamma]$ is positive for all positive $\gamma.$
This generates a more geometric transformation of fields, which is
\begin{equation} \label{eq:smeared-push}
	[\text{ANEC}[\gamma], \phi(x_+, x_-=0, \vec{y})]
	\sim \gamma(\vec{y}) \del_{x_+} \phi(x_+, x_-=0, \vec{y}).
\end{equation}
We can think of this transformation as pushing all fields located on the Rindler horizon forward along the horizon generators by an amount weighted by $\gamma$; see figure \ref{fig:null-push}.
While this transformation will generally act nonlocally for fields that are not localized on the Rindler horizon, causality implies that it must map operators in the Rindler wedge to operators in one of the ``null-deformed'' Rindler wedges sketched in figure \ref{fig:null-deformed-rindler-R}.
Let $\R$ be the Rindler wedge and $\R_{\gamma,s}$ be the wedge obtained by flowing for parameter value $s$ along the vector field specified by $\gamma.$
In the algebraic language, flowing by ANEC$[\gamma]$ with parameter $s$ generates a map from $\A(\R)$ to $\A(\R_{\gamma, s}).$

If ANEC$[\gamma]$ has positive expectation value in the state $|\Psi\rangle$, then that expectation value must tell us \textit{something} about how the state $|\Psi\rangle$ responds infinitesimally to the deformation of the algebra $\A(\R)$ towards the algebra $\A(\R_{\gamma,s}).$
One productive way to try to prove the ANEC is to guess what that \textit{something} is.
If we can show that the expectation value of the ANEC operator can be interpreted in terms of something that is manifestly positive, then we will have succeeded in proving the ANEC.

At this point, you could try a lot of different guesses.
I will tell you about the one that worked.
It has to do with an extremely natural quantity from the perspective of the algebraic approach and information theory, which is called \textit{relative entropy}.
In finite-dimensional quantum information theory, the relative entropy of two density matrices $\rho$ and $\sigma$ is
\begin{equation} \label{eq:relative-entropy}
	S(\rho\Vert\sigma)
		\equiv \tr(\rho \log{\rho}) - \tr(\rho \log{\sigma}).
\end{equation}
It is a quantum generalization of the Kullback-Leibler divergence of classical probability theory.
It is interpreted as a distinguishability measure; it is provably always positive, and its magnitude is interpreted as telling you how ``different'' $\rho$ is from $\sigma.$
There are of course many notions of ``distinguishability;'' for more on the specific interpretation of distinguishability due to relative entropy, see \cite{Witten:info-theory}.
One thing that is useful to know is that the relative entropy is an extremely sensitive measure of distinguishability; in fact, if $\rho$ and $\sigma$ are both pure and inequal, then the relative entropy between them is infinite.

One key feature of the relative entropy --- and one that validates its interpretation as a measure of distinguishability --- is that it is nonincreasing under partial trace.
If $AB$ is a bipartite system carrying density matrices $\rho_{AB}$ and $\sigma_{AB},$ and $\rho_{A}$ and $\sigma_{A}$ are the reduced density matrices on system $A$, then we have
\begin{equation} \label{eq:monotonicity}
	S(\rho_{AB}\Vert \sigma_{AB})
		\geq S(\rho_{A}\Vert \rho_B)
\end{equation}
This is called the monotonicity of relative entropy; it was first proved in \cite{Lieb:SSA}.

Since we will ultimately be thinking about relative entropy in quantum field theory, it is interesting to know that relative entropy can be defined rigorously for systems that do not admit density matrices.
This construction is due to Araki \cite{araki1975relative}.
For two states $|\Psi\rangle$ and $|\Phi\rangle$ on a Hilbert space that are cyclic and separating for a von Neumann algebra $\A$, there is a rigorous procedure that can be used to define a quantity $S_{\A}(|\Psi\rangle \Vert |\Phi\rangle)$ that reduces to equation \eqref{eq:relative-entropy} when it is possible to define density matrices for the subsystem described by the algebra $\A$.
For nested von Neumann algebras $\A\subseteq \A\B,$ a monotonicity relationship like \eqref{eq:monotonicity} holds.
For details on this construction, see \cite{Witten:entanglement}.

Let us now return to the Rindler wedge $\R$ and the sub-wedge $\R_{\gamma, s}$ sketched in figure \ref{fig:null-deformed-rindler-R}.
For any pair of states $|\Psi_1\rangle$ and $|\Psi_2\rangle,$ our ability to distinguish these states given access to $\R_{\gamma, s}$ is weaker than our ability to distinguish these states given access to $\R.$
If we choose a simple reference state --- for example, the vacuum $|\Omega\rangle$ --- then we should have the inequality
\begin{equation}
	S_{\R}(|\Psi\rangle \Vert |\Omega\rangle)
		\geq S_{\R_{\gamma, s}}(|\Psi\rangle \Vert |\Omega \rangle).
\end{equation}
Infinitesimally, this gives us the identity
\begin{equation}
	\del_s S_{\R_{\gamma,s}}(|\Psi\rangle \Vert |\Omega\rangle)|_{s=0} \leq 0.
\end{equation}
From the perspective of information theory, this is an extremely natural quantity associated with the transformation that takes $\R$ to $\R_{\gamma, s}$; it is the infinitesimal change in distinguishability between $|\Psi\rangle$ and the vacuum as we erase a little bit of the Rindler wedge $\R$.
Does it have anything to do with the ANEC operator ANEC$[\gamma]$?

\begin{figure}
	\centering
	\includegraphics{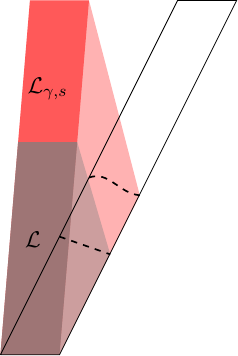}
	\caption{For the left Rindler wedge $\L$, deforming by ANEC$[\gamma]$ for parameter $s$ causes $\L$ to expand to a larger region $\L_{\gamma,s}$.}
	\label{fig:null-deformed-rindler-L}
\end{figure}

The answer is yes, it does; but the connection is not immediate.
The reason for this is that the ANEC operator does not just transform $\R$ to $\R_{\gamma,s}$, it also transforms the complementary wedge $\L$ to a larger wedge that we might call $\L_{\gamma,s}$; see figure \ref{fig:null-deformed-rindler-L}.
We also have the identity
\begin{equation}
	\del_{s} S_{\L_{\gamma,s}}(|\Psi\rangle \Vert |\Omega\rangle)|_{s=0} \geq 0.
\end{equation}
The quantity we will study is the sign-summed combination of these two derivatives:
\begin{equation} \label{eq:rel-ent-derivative}
	\del_s \left[S_{\L_{\gamma,s}}(|\Psi\rangle\Vert|\Omega\rangle) - S_{\R_{\gamma,s}}(|\Psi\rangle \Vert |\Omega \rangle)\right]|_{s=0} \geq 0.
\end{equation}
The key contribution of \cite{Faulkner:ANEC} was to argue that the LHS of the above equation is proportional to the expectation value of ANEC$[\gamma]$ in the state $|\Psi\rangle.$

If we write everything in terms of density matrices, the LHS of equation \eqref{eq:rel-ent-derivative} can be simplified to\footnote{This uses the purity relation $\tr(\chi_{A} \log \chi_{A}) = \tr(\chi_{B} \log \chi_B)$ for a state $\chi$ that is pure on the system $AB.$}
\begin{equation}
	\del_s \langle \Psi | \left( - \log \Omega_{\L_{\gamma,s}} + \log \Omega_{\R_{\gamma,s}} \right) | \Psi \rangle|_{s=0} \geq 0.
\end{equation}
This is the derivative of the expectation value in $|\Psi\rangle$ of what we called, in \hyperref[sec:lecture-1]{lecture one}, the \textit{full modular Hamiltonian} of $|\Omega\rangle$ with respect to the subsystem $\L_{s \gamma}.$
We may rewrite this as
\begin{equation}
	\del_s \langle \Psi | K_{\Omega, \L_{\gamma,s}}| \Psi \rangle|_{s=0} \geq 0.
\end{equation}
Indeed, one can show that full modular Hamiltonians for nested algebras always satisfy an operator monotonicity relation that reproduces this equation; for a rigorous proof of this fact that makes no reference to density matrices, see e.g. \cite{Witten:entanglement}.

\subsection{The algebraic argument for the ANEC}

In the previous subsection, we saw that we could prove the ANEC if we could argue for the identity
\begin{equation}
	\del_s K_{\Omega, \L_{\gamma,s}}|_{s=0}
		\propto \text{ANEC}[\gamma].
\end{equation}
For $s=0,$ we already know what the modular Hamiltonian $K_{\Omega, \L}$ is; we computed it in \hyperref[sec:lecture-1]{lecture one} as the appropriately normalized boost operator that is future-directed in the Rindler wedge $\L$.
So to compute the derivative in the above equation, we just need to understand the modular Hamiltonian $K_{\Omega, \L_{\gamma,s}}$ perturbatively to first order in $s.$
While there has not yet been produced a rigorous computation of this operator\footnote{See again footnote \ref{foot:rigor} for comments on rigor.} along the lines of the Bisognano-Wichmann computation \cite{Bisognano:1, Bisognano:2} of the modular Hamiltonian for $\L$, in \cite{Faulkner:ANEC} Faulkner, Leigh, Parrikar, and Wang performed a path integral manipulation that gives compelling evidence for the following form:\footnote{In fact, a beautiful argument from \cite{Casini:ANEC} shows that if this equation is true perturbatively, then it is exact --- if the order-$s$ term is the one claimed here, then there are no higher-order terms in $s$.}
\begin{equation} \label{eq:first-order-modham}
	K_{\Omega, \L_{\gamma,s}}
		= - 2 \pi \int dx_+\, d\vec{y}\, (x_+ - s \gamma(\vec{y})) T_{++}(x_+, x_-=0, \vec{y}) + O(s^2).
\end{equation}
In the limit $s\to0,$ this can be seen to reproduce the boost generator from equation \eqref{eq:boost}, after we switch the roles of $\L$ and $\R$ and use conservation of stress-energy to rewrite the boost generator as an integral over the surface $x_-=0.$
If we trust equation \eqref{eq:first-order-modham}, then it is easy to see
\begin{equation}
	\del_{s} K_{\Omega, \L_{\gamma,s}}|_{s=0}
		= 2 \pi \text{ANEC}[\gamma],
\end{equation}
so from equation \eqref{eq:first-order-modham}, the ANEC follows.

The argument for equation \eqref{eq:first-order-modham} is very much along the lines of the path-integral argument for the $s=0$ case given in \hyperref[sec:lecture-1]{lecture one}.
If we interpret equation \eqref{eq:first-order-modham} in terms of density matrices, then we see what we are trying to prove is
\begin{equation}
	- \log \Omega_{\L_{\gamma,s}}
		= \text{const} - 2 \pi \int_{x_+ \leq s \gamma(\vec{y})} dx_+\, d\vec{y}\, (x_+ - s \gamma(\vec{y})) T_{++}(x_+, x_-=0, \vec{y}) + O(s^2).
\end{equation}
If we can show this, then the full equation will follow from a symmetric argument.

\begin{figure}
	\centering
	\includegraphics{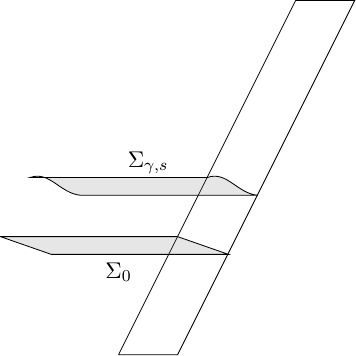}
	\caption{A cross-section $\Sigma_{\gamma, s}$ of the null-expanded Rindler wedge $\L_{\gamma, s}$.}
	\label{fig:null-deformed-rindler-cross-section}
\end{figure}

\begin{figure}
	\centering
	\includegraphics{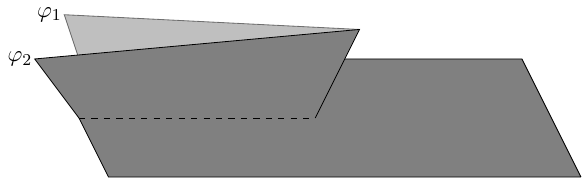}
	\caption{The complex manifold for which the path integral computes the matrix element $\langle \varphi_2 | \Omega_{\L_{\gamma, s}} |\varphi_1\rangle$.
	The ``flat'' portion of the manifold represents the Euclidean space $\reals^{d+1}$ that prepares the ground state; the ``raised'' or ``sewn-in'' portion represents the Lorentzian piece that evolves $\Omega_{\L}$ to $\Omega_{\L_{\gamma, s}}$.
	The $\vec{y}$ directions are not drawn; there are wiggles in these directions on the raised, sewn-in piece of the manifold.}
	\label{fig:complex-manifold}
\end{figure}

We can think of the density matrix $\Omega_{\L_{\gamma,s}}$ as an operator whose matrix elements are determined by field configurations on the surface $\Sigma_{\gamma,s}$ sketched in figure \ref{fig:null-deformed-rindler-cross-section}.
For two field configurations $\varphi_1$ and $\varphi_2$ on this surface, we would like to understand the matrix element
\begin{equation}
	\langle \varphi_2 | \Omega_{\L_{\gamma,s}} |\varphi_1\rangle.
\end{equation}
This is computed by a path integral over a complex metric $g$; there is a Euclidean piece that prepares the ground state, then a Lorentzian piece that evolves from $\Sigma_{0}$ to $\Sigma_{\gamma,s}.$
The complex manifold over which the path integral is performed is sketched in figure \ref{fig:complex-manifold}.
By picking a diffeomorphism that maps this complex manifold to Euclidean space, we can express matrix elements of $\Omega_{\L_{\gamma,s}}$ in terms of matrix elements of $\Omega_{\L}$.

To be very concrete, let us denote the complex manifold by $\M$ with metric $g.$
We will be mapping to the Euclidean manifold $\mathbb{R}^{d+1}$ with standard flat metric $\eta.$
The path integral we want to compute is
\begin{equation}
	\langle \varphi_2 | \Omega_{\L_{\gamma,s}} |\varphi_1\rangle
		= \frac{1}{Z} \int_{\phi|_{\Sigma_{s \gamma},+} = \varphi_1}^{\phi|_{\Sigma_{s \gamma},-} = \varphi_2} \D_{\M, g} \phi e^{I_{\M, g}[\phi]},
\end{equation}
where $I$ is the appropriate complexified action.
Given a diffeomorphism $\psi : \M \to \mathbb{R}^{d+1},$ this path integral is identically equal to
\begin{equation}
	\langle \varphi_2 | \Omega_{\L_{\gamma,s}} |\varphi_1\rangle
		= \frac{1}{Z} \int_{\phi|_{\Sigma_{0},+} = \varphi_1 \circ \psi^{-1}}^{\phi|_{\Sigma_{0},-} = \varphi_2 \circ \psi^{-1}} \D_{\mathbb{R}^{d+1}, \psi^* g} \phi e^{I_{\mathbb{R}^{d+1}, \psi^*g}[\phi]}.
\end{equation}
If we are judicious in our $s$-dependent choice of diffeomorphism $\psi$, we will be able to write $\psi^* g = \eta + s\, \delta \eta + O(s^2).$
The linear response of the path integral measure to a change in the metric is controlled by the stress tensor, and if you are sufficiently careful about minus signs and factors of $i$, you will find that the resulting expression is 
\begin{equation}
	\langle \varphi_2 | \Omega_{\L_{\gamma,s}} |\varphi_1\rangle
	= \frac{1}{Z} \int_{\phi|_{\Sigma_{0},+} = \varphi_1 \circ \psi^{-1}}^{\phi|_{\Sigma_{0},-} = \varphi_2 \circ \psi^{-1}} \D_{\mathbb{R}^{d+1}, \eta} \phi e^{I_{\mathbb{R}^{d+1}, \eta}[\phi]} \left( 1 - s\, \frac{i}{2} \int d^{d+1}x\, T^{(\eta)}_{\mu \nu} \delta \eta^{\mu \nu} + O(s^2) \right).
\end{equation}

\begin{figure}
	\centering
	\includegraphics{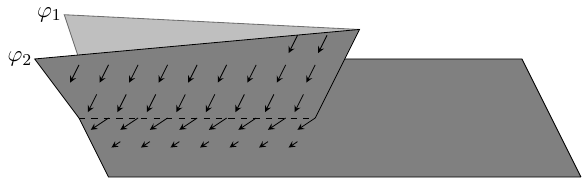}
	\caption{A vector field $X^{\mu}$ that, when integrated, compresses away the ``sewn-in'' piece of the manifold from figure \ref{fig:complex-manifold}.}
	\label{fig:complex-vector-field}
\end{figure}

To construct our family of diffeomorphisms $\psi$, we recall from figure \ref{fig:complex-manifold} that $\M$ consists of a Euclidean piece that produces the vacuum state, together with a ``sewn in'' piece that has a Lorentzian metric.
We will construct $\psi$ by choosing a vector field $X^{\mu}$ that vanishes away from a neighborhood of this ``sewn in'' piece, and that compresses the sewn-in piece into the flat manifold $\reals^{d+1}$ after integrating by parameter $s$; the diffeomorphism $\psi$ is obtained by integrating this vector field.
For small $s,$ we may choose
\begin{equation}
	X^{\mu} = F(\tau, x, \vec{y}) \left[ - \gamma(\vec{y}) \left[ \left( \frac{\del}{\del \tau} \right)^{\mu} + \left( \frac{\del}{\del x} \right)^{\mu} \right] + O(s) \right],
\end{equation}
where $F$ is a function that equals one in a neighborhood of the piece that was sewn in, and vanishes away from a neighborhood of that piece.
See figure \ref{fig:complex-vector-field}.
From this, we have
\begin{equation}
	\delta \eta^{\mu \nu} = \del^{\mu} X^{\nu} + \del^{\nu} X^{\mu}.
\end{equation}
Plugging this into our above expression and exploiting the symmetry of the stress tensor gives
\begin{equation} \label{eq:deformed-matrix-elements}
	\langle \varphi_2 | \Omega_{\L_{\gamma,s}} |\varphi_1\rangle
	= \frac{1}{Z} \int_{\phi|_{\Sigma_{0},+} = \varphi_1 \circ \psi^{-1}}^{\phi|_{\Sigma_{0},-} = \varphi_2 \circ \psi^{-1}} \D_{\mathbb{R}^{d+1}, \eta} \phi e^{I_{\mathbb{R}^{d+1}, \eta}[\phi]} \left( 1 - i s \int d^{d+1}x\, T^{(\eta)}_{\mu \nu} \del^{\mu} X^{\nu} + O(s^2) \right).
\end{equation}

The rest of the argument is very subtle, and entails a detailed manipulation of equation \eqref{eq:deformed-matrix-elements} that is both technical and heuristic.
Essentially, the right-hand side of equation \eqref{eq:deformed-matrix-elements} is interpreted as computing matrix elements of some complicated operator $W_{\gamma,s}$ on the Hilbert space of field configurations on $\Sigma_{\gamma,0}$:
\begin{equation}
	\langle \varphi_2| \Omega_{\gamma,s} | \varphi_1 \rangle
		= \langle \varphi_2 \circ \psi^{-1} | W_{\gamma,s} | \varphi_1 \circ \psi^{-1} \rangle.
\end{equation}
Considering the diffeomorphism $\psi$ as a unitary map that exchanges the Hilbert space of field configurations on $\Sigma_{\gamma,s}$ for the Hilbert space of field configurations on $\Sigma_{\gamma,0},$ one obtains the following expression for the logarithm:
\begin{equation} \label{eq:log-matrix-elements}
	\langle \varphi_2 | \log\Omega_{\gamma,s} |\varphi_1\rangle
	= \langle \varphi_2 \circ \psi^{-1} | \log W_{\gamma,s} | \varphi_1 \circ \psi^{-1} \rangle.
\end{equation}
At this point, one observes that the operator $W_{\gamma,s}$ is related to the Rindler density matrix $\Omega_{\L}$ by
\begin{equation}
	W_{\gamma,s}
		= \frac{1}{Z'} \left[ \Omega_{\L} - i s \mathcal{T}\left\{ \Omega_{\L} \int d^{d+1} x\, T_{\mu \nu}^{\eta} \del^{\mu} X^{\nu} \right\} + O(s^2) \right],
\end{equation}
where $\mathcal{T}$ denotes an angle-ordered product.
This is a somewhat fishy expression, as there are subtleties involved in interpreting an angle-ordered product as an operator acting on a Hilbert space, rather than as a formal expression that makes sense within correlation functions.
If however you trudge ahead and interpret it as an operator, then you can plug it into the integral identity\footnote{See \cite[section 2]{Balakrishnan:2020lbp} for a clear derivation of this identity and a nice generalization of the main calculation performed in \cite{Faulkner:ANEC}.}
\begin{equation} \label{eq:logarithm-expansion}
	\log(A + s B)
		= \log{A} - s \lim_{\epsilon \to 0^+} \int_{-\infty}^{\infty} d\lambda\, \frac{1}{4 \sinh\left(\frac{\lambda+i \epsilon}{2}\right)^2} A^{i \lambda/2\pi} A^{-1} B A^{-i\lambda/2\pi}  + O(s^2).
\end{equation}
The operator ``$A$'' in this expression will be the vacuum density matrix $\Omega_{\L_0}$, so the terms $A^{\pm i\lambda/2\pi}$ will implement geometric boosts.
The operator ``$B$'' in this expression will be some complicated time-ordered product of boosts and stress tensor insertions.
In \cite{Faulkner:ANEC}, formal manipulations of this expression showed that
\begin{enumerate}[(i)]
	\item Consistent treatment of the path integral near the bifurcation surface $\tau=x=0$ requires introducing an ultraviolet cutoff in the path integral; they choose to excise a small portion of the Euclidean manifold around the bifurcation surface.
	\item Contributions to the stress-tensor integral away from the cutoff surface cancel with the contribution to equation \eqref{eq:log-matrix-elements} coming from the change in boundary conditions due to the diffeomorphism $\psi.$
	\item A boundary term localized on the cutoff surface does not vanish, and the integral over boosts in equation \eqref{eq:logarithm-expansion} ends up smearing this contribution over a boost trajectory whose distance from $x=\tau=0$ is set by the cutoff scale.
	\item In the limit as the cutoff is removed, the ``boost-smeared'' stress-tensor term limits to the appropriate half-ANEC operator needed to reproduce equation \eqref{eq:first-order-modham}.
\end{enumerate}

\section{Lecture 3: Semiclassical black holes}
\label{sec:lecture-3}

\subsection{Semiclassical black hole entropy}

For this lecture we will switch gears and talk about a very recent development in high energy physics coming from the algebraic approach.
This is a development that has to do with understanding the entropy of black holes in semiclassical quantum gravity.
The main relevant papers are \cite{Witten:crossed, Chandrasekaran:dS, Akers:entropy}.
Some other relevant work includes \cite{Chandrasekaran:BH, Penington:JT, Kolchmeyer:JT, Jensen:gen-ent, Kudler-Flam:BH}.

It was realized in the 1970s that classical black holes in general relativity obey laws that are very similar to the thermodynamic laws satisfied by finite-temperature matter systems \cite{Bardeen:1973gs, Bekenstein:1973ur}, provided that one interprets area as entropy, mass as energy, and surface gravity as temperature.
Thanks to the pioneering work of Boltzmann \cite{Boltzmann} and Gibbs \cite{Gibbs}, it is known that the laws of thermodynamics for ordinary matter systems can be reproduced from an underlying statistical description of a many-body system with a large number of microstates.
This led to a great deal of speculation about the quantum nature of black holes --- are the laws of black hole mechanics also derivable from statistical mechanics?
Does the area of a black hole's event horizon measure the fine-grained entropy of an underlying ensemble of states in quantum gravity?

With the advent of D-brane physics, this question was answered positively for certain classes of black holes in string theory; see e.g. \cite{Strominger:strings, Callan:micro, Horowitz:micro, Maldacena:micro, Johnson:micro}.
These papers constructed large families of string theory microstates consistent with macroscopic black hole parameters, and showed that in an appropriate semiclassical limit, the number of microstates scales like the exponential of the black hole area in Planck units.
Other work using holographic theories of quantum gravity has established that similar behavior can be found without needing to invoke string theory explicitly \cite{Strominger:CFT, Hartman:light-spectrum, Mukhametzhanov:tauberian, Balasubramanian:2022gmo}.

There is an interesting gap in the literature we have described so far.
We have described the ``thermodynamic'' laws of black holes, and we have described the ``quantum statistical mechanical'' microstate counting of certain black holes in string theory.
But the Boltzmann-Gibbs theory of statistical mechanics, which reproduced the empirical laws of thermodynamics, was not quantum!
Boltzmann and Gibbs never actually counted microstates of thermodynamic systems, since the number of classical microstates of any thermodynamic system is infinite.
Instead, they defined \textit{ratios} of numbers of microstates --- they came up with a way to make statements like ``there are $N$ times as many classical microstates for configuration $A$ as for configuration $B$,'' and inferred that system $A$ was statistically preferred to system $B$, even though the number of microstates consistent with each configuration was formally infinite.
The desire to resolve these infinities then became one of the main motivations for the development of quantum mechanics.

The question we now ask is: does the entropy of semiclassical black holes admit a well defined ``relative'' description --- like the description of black hole entropy in classical statistical mechanics --- without reference to a specific theory of quantum gravity such as string theory?
Such a description is obviously desirable because (i) it would apply to every semiclassical black hole, not just those that can be explicitly constructed in string theory, and (ii) it would provide helpful guideposts for further development of microscopic quantum gravity.

The algebraic approach is extremely well suited for posing and answering this question.
What we are really asking is: can the entropy of a black hole be interpreted as a measure of the information available to semiclassical observables?
The semiclassical observables in the vicinity of a black hole form an algebra, and what we are really asking is, ``is there a consistent and useful way to talk about entropy in reference to this algebra?''
The answer is yes.

\subsection{Traces, types, and entropy}

A major development in our understanding of semiclassical black hole entropy was made by Witten in \cite{Witten:crossed}.
The main idea --- inspired by earlier investigations in \cite{Segal:entropy} --- was to temporarily throw away physical interpretation, and begin with a mathematical declaration: an \textit{entropy} is a quantity of the form
\begin{equation}
	- \tr(\rho \log \rho).
\end{equation}
So in order to be able to define an entropy for a system, we need two things: a trace operation $\tr$, and a density matrix $\rho.$
Let us be a little more explicit.
Suppose we have a von Neumann algebra $\A$ of operators acting on a Hilbert space $\H$, and a state $|\Psi\rangle$ on that Hilbert space.
The Hilbert space comes with a natural trace operation on $\A$.
We say that there is a \textit{density matrix} for $|\Psi\rangle$ if there is a positive operator $\rho \in \A$ with
\begin{equation}
	\tr(\rho a)
		= \langle \Psi | a | \Psi \rangle, \qquad a \in \A.
\end{equation}
The entropy of the state $|\Psi\rangle$ in the subsystem described by the algebra $\A$ is then defined by $- \tr(\rho \log \rho).$

It is well known --- see e.g. \cite{Fredenhagen:1984dc} --- that for the von Neumann algebras describing subregions in quantum field theory, there are no density matrices for any state.
In fact, for a spacetime region $\O$, there are generally no operators in $\A(\O)$ with finite trace!
This means there can be no density matrices in $\A(\O)$, since these would need to have $\tr(\rho) = 1.$
One might imagine, however, that it is possible to define a functional $\tau$ on the algebra $\A(\O)$ that plays a similar role to the Hilbert space trace, and then to define density matrices and entropies with respect to this ``modified'' trace.
This idea is at the core of the mathematical type classification of von Neumann algebras \cite{murray1936rings, murray1937rings, neumann1940rings, murray1943rings}, where von Neumann algebras are classified by whether or not it is possible to define a modified trace, and by the properties of the modified trace functional.
In the context of physics, a key step forward was made when Witten in \cite{Witten:crossed} observed that the modified traces that show up in the type classification of von Neumann algebras can be interpreted in terms of physical renormalization schemes; this perspective was then further developed in \cite{Sorce:notes}.

The technical definition of a ``modified trace'' --- which we will now start to call a ``renormalized trace'' --- is given below.
\begin{definition}
	Given a von Neumann algebra $\A$ on the Hilbert space $\H$, we denote by $\A_+$ the space of positive operators.
	A \textbf{trace} is a map $\tau : \A_+ \to [0, \infty]$ satisfying linearity:
	\begin{equation}
		\tau(\lambda(a + b)) = \lambda \tau(a) + \lambda \tau(b) \qquad a, b \in \A_+, \lambda \geq 0
	\end{equation}
	and unitary invariance:
	\begin{equation}
		\tau(U a U^{\dagger}) = \tau(a), \qquad a \in \A_+, U \in \A \text{ unitary}.
	\end{equation}
	$\tau$ is a \textbf{renormalized trace} if it satisfies the following three natural physical properties:
	\begin{itemize}
		\item Faithful: If $\tau(a)$ vanishes then $a$ vanishes.
		\item Normal: If $a_1 \leq a_2 \leq \dots \leq a$ is a sequence of positive operators in $\A_+$ with $\sup_{n} a_n = a$, then we have $\tau(a_n) \to \tau(a)$.
		(This is just a version of continuity that works for functionals that can take the value infinity.)
		\item Semifinite: The set of operators $a \in \A$ with $\tau(|a|) < \infty$ is dense in $\A$ in an appropriate topology.
		(Typically the \textit{ultraweak} topology, though there are a few choices that give the same answer.)
	\end{itemize}
\end{definition}
\begin{remark}
	From this definition, it is possible to show many useful properties of renormalized traces; see for example \cite[appendix B]{Sorce:notes}.
	The most important property is that the space of operators in $\A$ with $\tau(|a|) < \infty$ forms a subspace of $\A$ on which $\tau$ can be defined consistently as a linear, cyclic functional.
	In fact, this subspace is an ideal, meaning if you take an operator $a$ with $\tau(|a|) < \infty,$ then for any $b \in \A$, we also have $\tau(|ab|) < \infty$ and $\tau(|ba|) < \infty.$
	A density matrix $\rho$ is simply a positive member of this ideal with $\tau(\rho) = 1$; because of the ideal property, all expectation values $\tau(\rho a)$ are well defined for any $a \in \A.$
\end{remark}

The concept of a renormalized trace is used to classify von Neumann algebras into types.
For details, see \cite{Sorce:notes}.
The basic idea is first to show that every von Neumann algebra can be decomposed into a direct sum of ``factors'' --- these are the algebras that typically show up in physics, and they are defined by the property that the only elements of $\A$ that commute with everything in $\A$ are the multiples of the identity.
One then shows that on a factor, if a renormalized trace exists, it is ambiguous only up to an overall scale; in other words, on a factor $\A$, any two renormalized traces $\tau_1$ and $\tau_2$ must satisfy $\tau_1 = c \tau_2$ for some positive constant $c.$
The classification of factors is then as follows.
\begin{definition}
	\begin{itemize}
	\item A factor $\A$ is of \textbf{type III} if it admits no renormalized trace.
	\item A factor $\A$ is of \textbf{type II} if it admits a unique one-parameter family of renormalized traces, and if the traces of nonzero projectors in $\A$ can be arbitrarily close to zero.
	\item A factor $\A$ is of \textbf{type I} if it admits a unique one-parameter family of renormalized traces, and if the traces of nonzero projectors in $\A$ have a minimal value.
	\end{itemize}
\end{definition}
Type III factors are ones for which no renormalized entropy can be defined.
Type I factors are ones for which a renormalized entropy can be defined, and for which there are ``pure states'' --- minimal projection operators with respect to the trace --- that can be used to canonically normalize the trace.
Type II factors are ones for which a renormalized entropy can be defined, but for which the trace cannot be canonically normalized.
In type II factors, the lingering multiplicative ambiguity in the trace leads to an additive ambiguity in the entropy $- \tau(\rho \log \rho).$
This is analogous to the situation in classical statistical mechanics; we will now see that this mathematical structure arises naturally in the context of semiclassical black holes.

\subsection{The crossed product}

It is generally expected, thanks to the work of \cite{Fredenhagen:1984dc}, that in quantum field theory local field algebras are of type III.
This means that density matrices and entropies cannot be defined even in a renormalized way.
In \cite{Witten:crossed}, Witten proposed that coupling gravity to quantum field theory in the vicinity of a black hole leads to an algebra of observables that is type II, meaning that for semiclassical gravity, entropies of states are well defined up to an additive ambiguity --- consequently, entropy differences can be unambiguously defined as in classical statistical mechanics.
In \cite{Chandrasekaran:dS} (see also \cite{Jensen:gen-ent}), it was shown that the entropy differences so-defined are consistent with the Bekenstein-Hawking formula for semiclassical black hole entropy.

The key tool is a mathematical construction that was developed by Takesaki in \cite{Takesaki1973} in an effort to understand the mathematical properties of type III von Neumann algebras.
This is called the \textit{crossed product}; it nicely incorporates both the type structure we have discussed in this lecture so far, and the modular flow that we discussed in \hyperref[sec:lecture-1]{lecture one}.
A detailed, pedagogical explanation of the crossed product can be found in \cite[appendix B]{Jensen:gen-ent}.

Mathematically, the idea is very simple.
Given a type III von Neumann algebra $\A$ acting on the Hilbert space $\H$, and given a cyclic-separating state $|\Psi\rangle$ with modular Hamiltonian $K_{\Psi},$ Takesaki proposed enlarging $\A$ into an algebra that explicitly incorporates the corresponding modular flow.
What he does is tack on an extra Hilbert space $L^2(\reals)$, and define a von Neumann algebra on $\H \otimes L^2(\reals)$ with the following generating set:
\begin{equation}
	\A \rtimes K_{\Psi}
		\equiv \langle a \otimes 1, e^{i (K_{\Psi} + \hat{p}) s}\, |\, a \in \A, s \in \mathbb{R}\rangle.
\end{equation}
The operator $e^{i(K_{\Psi} + \hat{p})s}$ is interpreted as implementing modular flow on elements of $\A$, but keeping track of the amount of modular flow that has been performed by doing a simultaneous translation of the auxiliary register $L^2(\mathbb{R}).$
Takesaki showed that if $\A$ is of type III,\footnote{Really it needs to be a special kind of type III algebra called a type III$_1$ algebra, but these are typical in quantum field theory.
See \cite{Sorce:notes}.} then $\A \rtimes K_{\Psi}$ is of type II.

So far this is an interesting piece of mathematical trivia, but it is not so easy to see how it is connected to physics.
The connection is provided by the \textit{commutation theorem} \cite{van1978continuous}, which gives a unitarily equivalent description of the crossed product algebra as
\begin{equation} \label{eq:commutation-theorem}
	\A \rtimes K_{\Psi}
		\cong \{ \tilde{a} \in \A \otimes \B(L^2(\mathbb{R}))\, | [K_{\Psi} - \hat{x}, \tilde{a}] = 0 \}.
\end{equation}
So Takesaki's theorem has the following interpretation: if you take a type III algebra $\A$, enhance it by adding in all possible operators on an auxiliary space L$^2(\mathbb{R}),$ then restrict to those operators that commute with $K_{\Psi} - \hat{x},$ you get a type II algebra on which entropy difference can be defined.

This mathematical structure shows up very naturally in black hole physics.
In classical gravity, the exterior of a Schwarzschild black hole carries a type III algebra of quantum fields, $\A_{\text{QFT}},$ acting on a Hilbert space $\H_{\text{QFT}}.$
When gravity is quantized perturbatively, it is necessary to add an additional Hilbert space $\H_{\text{grav}}$ of linearized metric excitations around the black hole background.
Within this Hilbert space is a subspace, $\H_{\text{mass}},$ which describes excitations that only change the mass of the black hole, without changing any of its other parameters.
This Hilbert space can be identified with $L^2(\mathbb{R}),$ since its elements are wavefunctions of mass fluctuation.
The equations of general relativity enforce a relationship between the state of the matter fields and the magnitude of the mass fluctuation; in a Schwarzschild black hole, this constraint is
\begin{equation}
	\hat{M}_{\text{BH}} = \int_{t=0} T_{\mu \nu} \left( \frac{\del}{\del t} \right)^{\mu}  d \Sigma^{\nu}, 
\end{equation}
with $t$ the Schwarzschild time coordinate.
The naive observable algebra $\A_{\text{QFT}} \otimes \B(\H_{\text{mass}})$ does not know about this constraint --- the \textit{physical} operators are those that commute with the operator $ \int_{t=0} T_{\mu \nu} \left( \frac{\del}{\del t} \right)^{\mu}  d \Sigma^{\nu} - \hat{M}_{\text{BH}}.$
The algebra of physical operators in the black hole exterior is therefore
\begin{equation} \label{eq:commutator-constraint}
	\A_{\text{phys}}
		= \left\langle \tilde{a} \in \A_{\text{QFT}} \otimes \B(\H_{\text{mass}})\, | \left[\int_{t=0} T_{\mu \nu} \left( \frac{\del}{\del t} \right)^{\mu}  d \Sigma^{\nu} - \hat{M}_{\text{BH}}, \tilde{a}\right] = 0 \right\rangle.
\end{equation}
This is extremely similar to equation \eqref{eq:commutation-theorem}; the operator $\hat{M}_{\text{BH}}$ on $\H_{\text{mass}}$ indeed plays exactly the same role as the position operator on $L^2(\mathbb{R}).$
The key difference is that in equation \eqref{eq:commutation-theorem} we had a modular Hamiltonian, while in equation \eqref{eq:commutator-constraint} we have an integral of the stress tensor.
In fact, it is the integral of the stress tensor that generates Schwarzschild time translations.

Hang on --- that looks pretty familiar!
In \hyperref[sec:lecture-1]{lecture one} we showed that the Minkowski vacuum restricted to a Rindler wedge has as its modular Hamiltonian $2\pi$ times the integral of the stress tensor that generates boosts in the Rindler wedge.
Indeed, an identical argument shows that the Hartle-Hawking state $|\Psi_{\text{HH}}\rangle$ in the background of a Schwarzschild black hole has, as its modular Hamiltonian, the integral
\begin{equation}
	K_{\Psi_{\text{HH}}}
		= \frac{2 \pi}{\kappa} \int_{t=0} T_{\mu \nu} \left( \frac{\del}{\del t}\right)^{\mu} d\Sigma^{\nu},
\end{equation}
where $\kappa$ is the surface gravity of the black hole.
So we have
\begin{equation}
	\A \rtimes K_{\Psi_{\text{HH}}}
		\cong \left\langle \tilde{a} \in \A_{\text{QFT}} \otimes \B(\H_{\text{mass}})\, | \left[\frac{2 \pi}{\kappa} \int_{t=0} T_{\mu \nu} \left( \frac{\del}{\del t} \right)^{\mu}  d \Sigma^{\nu} - \hat{M}_{\text{BH}}, \tilde{a}\right] = 0 \right\rangle.
\end{equation}
We can clearly furnish an isomorphism between this algebra and $\A_{\text{phys}}$ by rescaling the mass by $2\pi/\kappa$; this shows that the algebra of physical operators in the black hole exterior, $\A_{\text{phys}},$ is of the type II kind that admits well defined differences in entropy.

In fact, the work of Takesaki \cite{Takesaki1973} gives us way more.
His paper not only shows that the crossed product algebra is type II, but constructs an explicit formula for a renormalized trace on the crossed product.
If you look at his formula and translate it into our language, you find that one member of the family of renormalized traces on $\A_{\text{phys}}$ is given by
\begin{equation}
	\tau(\tilde{a})
		= \left( \langle \Psi|_{\text{HH}} \otimes \langle 0|_{p} \right) e^{\frac{\kappa}{4 \pi} \hat{M}_{\text{BH}}} \tilde{a} e^{\frac{\kappa}{4 \pi} \hat{M}_{\text{BH}}}
			\left( | \Psi \rangle_{\text{HH}} \otimes | 0\rangle_{p} \right).
\end{equation}
In this expression, $|0\rangle_p$ is the unnormalizable momentum eigenstate for the ``momentum'' variable conjugate to $\hat{M}_{\text{BH}}$ on $\H_{\text{mass}}.$
This is a cool expression, because it is simple enough to be manipulated by anyone who knows single particle quantum mechanics.
For a semiclassical state of the black hole $|\Phi\rangle_{\text{QFT}} \otimes |f\rangle_{\text{mass}},$ one can seek a corresponding density matrix $\rho_{\Phi, f}$ satisfying the identity
\begin{equation}
	\tau(\rho_{\Phi, f} \tilde{a})
		= \left( \langle \Phi |_{\text{QFT}} \otimes \langle f|_{\text{mass}} \right) \tilde{a} \left( |\Phi\rangle_{\text{QFT}} \otimes |f\rangle_{\text{mass}} \right).
\end{equation}
for every $\tilde{a} \in \A_{\text{phys}}$.
An expression for $\rho_{\Phi, f}$ in a certain approximation was found in \cite{Chandrasekaran:dS}, and a non-approximate formula was produced in \cite{Jensen:gen-ent}.
These papers then computed the entropy $-\tau(\rho_{\Phi, f} \log \rho_{\Phi, f})$ in an approximation where the wavefunction $f$ is slowly varying --- this approximation was then rigorously justified in \cite{Kudler-Flam:approx} --- and obtained a completely well defined, UV-finite formula
\begin{equation}
	S(\rho_{\Phi, f})
		= - S_{\A}(\Phi_{\text{QFT}}\Vert \Psi_{\text{HH}}) + \frac{\kappa}{2\pi} \langle \hat{M}_{\text{BH}} \rangle_{f_{\text{mass}}} - \int_{-\infty}^{\infty} dx |f(x)|^2 \log |f(x)|^2 + \text{const}.
\end{equation}
The overall constant term cannot be removed, because it is related to the scaling ambiguity in the definition of $\tau$.
Note that this expression includes the relative entropy, which we explored in our study of the ANEC in \hyperref[sec:lecture-2]{lecture two}.
 
Because the additive ambiguity in the entropy formula is state-independent, the formula suffices to compute entropy differences.
It was shown in \cite{Chandrasekaran:dS, Jensen:gen-ent, AliAhmad:2023etg} that if this UV-finite formula is regulated in a natural way, then the gravitational equations of motion enforce that the entropy can be interpreted as the area of the black hole in Planck units, plus a matter entropy, plus a state-independent constant.
This observation is the source of considerable excitement around the algebraic approach to understanding black hole entropy --- the rigorously well defined UV-finite entropy differences in semiclassical black hole backgrounds is consistent with the Bekenstein-Hawking prediction for the entropy of a semiclassical black hole.

\subsection{A preliminary interpretation}

We will finish this lecture with a quick discussion of the recent results of \cite{Akers:entropy}.
So far, our approach to thinking about the entropy of semiclassical black holes has been purely formal --- we sought to define an expression $-\tau(\rho \log \rho)$ that \textit{looks like} an entropy, found that the answer was uniquely determined up to an additive constant, and then discovered, quite miraculously, that the answer is consistent with the Bekenstein-Hawking entropy formula.
But what right do we have to interpret $-\tau(\rho \log \rho)$ as an entropy?
Does it have a state-counting interpretation like the entropy in statistical mechanics?

First, let us think for a moment about entropy in classical statistical mechanics.
For any probability distribution $p(\lambda)$ on phase space, the entropy
\begin{equation}
	S = - \int d\lambda\, p(\lambda) \log p(\lambda)
\end{equation}
is well defined up to the additive constant coming from the necessity of choosing units on phase space.
But not every probability distribution has the property that the entropy can be interpreted as counting microstates.
The distributions for which the entropy has a microstate interpretation are the microcanonical distributions, where $p(\lambda)$ is uniform on a subset of phase space.
In other settings, the entropy might have an approximate interpretation in terms of counting states --- this underlies the ``smooth entropies'' defined in quantum information theory \cite{renner2004smooth, renner2008security} --- but the microcanonical ensembles are the ones for which the connection between entropy and microstates is most clear.

So what about the type II entropy $- \tau(\rho \log \rho)$?
First, what does it mean for $\rho$ to represent a microcanonical ensemble of black hole states?
Second, can the type II entropy difference of microcanonical ensembles be interpreted as a relative microstate count?

These questions were answered in \cite{Akers:entropy}.
Remember that $\rho$ must be a positive operator in $\A_{\text{QFT}} \otimes \B(\H_{\text{mass}})$ that is invariant under the action of $\frac{\kappa}{2 \pi} K_{\Psi}  - \hat{M}_{\text{BH}}.$
The density matrices that represent canonical ensembles of black hole microstates are ones of the form
\begin{equation}
	\rho \propto 1_{\text{QFT}} \otimes P_{\text{mass}},
\end{equation}
where $P_{\text{mass}}$ is a projection operator onto a compact window of black hole masses.
It is easy to verify that this operator is in the physical algebra, and that it has finite renormalized trace.
In \cite{Akers:entropy}, we used techniques from the mathematical theory of operator algebras to show the following.
\begin{theorem}
	Let $\rho$ and $\sigma$ be two density matrices in $\A_{\text{phys}}$ representing microcanonical ensembles of black hole mass.
	Let $\H_{\rho}$ and $\H_{\sigma}$ be the subspace of Hilbert space on which $\rho$ and $\sigma$ are supported, respectively.
	
	Suppose without loss of generality $\Delta S \equiv S(\rho) - S(\sigma) \geq 0.$
	Then for any integer $n \geq e^{\Delta S}$, $\H_{\rho}$ can be embedded into $\H_{\sigma} \otimes \comps^{n}$ using unitary operators in $\A_{\text{phys}}$; for any integer $n < e^{\Delta S},$ no such embedding exists.
\end{theorem}
The way we should think of this theorem is as telling us that $e^{\Delta S}$ gives the extra number of degrees of freedom that are needed to get from the portion of Hilbert space where $\sigma$ is supported to the portion of Hilbert space where $\rho$ is supported.
While both of these Hilbert spaces are infinite, the notion of the relative size of $\H_{\rho}$ as compared to $\H_{\sigma}$ is well defined in terms of the above theorem.
This is the sense in which the type II entropy difference computes a relative microstate count for semiclassical black holes.
It remains an open question whether this line of reasoning helps explain the classical laws of black hole thermodynamics, as the Boltzmann-Gibbs theory explained the laws of thermodynamics for matter.

\acknowledgments{
I am grateful to Onkar Parrikar for providing feedback on a draft of lecture two, and to Slava Rychkov and the other organizers of Bootstrap 2024 for inviting me to prepare these lectures.
I am also grateful to all of the attendees of Bootstrap 2024 for excellent questions that helped in revising these notes for future use.
All of my work is supported by the AFOSR under award number FA9550-19-1-0360, by the DOE Early Career Award number DE-SC0021886, and by the Heising-Simons Foundation.
}

\appendix

\section{Angular slicing of the path integral}
\label{app:angular-slicing}

Here we give a simple derivation of the statement that the matrix elements defined by figure \ref{fig:vacuum-density-matrix} are matrix elements of the expression
\begin{equation}
	\exp\left[ - 2 \pi \int_0^{\infty} dx\, x\, x T^{E}_{\tau \tau}\right],
\end{equation}
with $T^E$ the Euclidean stress-energy tensor and $\tau$ the Euclidean time coordinate.

\begin{figure}
	\centering
	\includegraphics[scale=1.5]{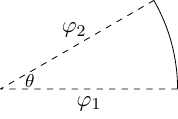}
	\caption{Given field configurations $\varphi_1$ and $\varphi_2,$ we map $\varphi_2$ to the slice $\Sigma_{\theta}$ in a natural way, and use the resulting path integral to define the matrix elements of an operator $W_{\theta}$.}
	\label{fig:wedge-path-integral}
\end{figure}

Let $\Sigma_\theta$ denote the codimension-1 surface in Euclidean space defined by the equations
\begin{align}
	\tau
		& = \sqrt{\tau^2 + x^2} \sin{\theta}, \\
	x
		& = \sqrt{\tau^2 + x^2} \cos{\theta}.
\end{align}
Since slices of constant $\theta$ can be mapped to one another by a preferred rotation symmetry, it is easy to identify the field configuration $\varphi$ on the surface $\Sigma_{\theta_1}$ with the field configuration $\varphi \circ R_{\theta_1 - \theta_2}$ on $\Sigma_{\theta_2}.$
Given two field configurations $\varphi_1$ and $\varphi_2$ on $\Sigma_0,$ it is therefore natural to consider the ``wedge'' path integral shown in figure \ref{fig:wedge-path-integral}; this is the path integral over over a portion of Euclidean space with angular extent $\theta$, with $\varphi_1$ the ``bottom'' boundary condition and $\varphi_2 \circ R_{-\theta}$ the ``top'' boundary condition.
We will call the outcome of this path integral $\langle \varphi_2 | W_{\theta} |\varphi_1\rangle$, and take this to define the operator $W_{\theta}$; the quantity we want to compute is $\langle \varphi_2 | W_{2 \pi} |\varphi_1\rangle.$
We will get there by computing its $\theta$ derivative; if we can show that the $\theta$ derivative satisfies
\begin{equation} \label{eq:appendix-desired-equation}
	\del_{\theta} \langle \varphi_2 | W_{\theta} | \varphi_1\rangle
		= - \langle \varphi_2 | K_{\theta} W_{\theta} | \varphi_1 \rangle,
\end{equation}
with $K_{\theta}$ defined in equation \eqref{eq:angle-generator}, then the discussion immediately following that equation leads to our desired result.

For any $\theta_1$ and $\theta_2$ with $\theta_1 < \theta_2,$ we can get a diffeomorphism that maps the $\theta_2$ wedge to the $\theta_1$ wedge by choosing a function $f(\theta)$ that is zero at $\theta=0$ and one in $\theta_1 \leq \theta \leq \theta_2,$ then integrating the vector field $- f(\theta) \frac{\del}{\del \theta}$ from zero to $\theta_2 - \theta_1.$
Call this diffeomorphism $\psi_{f, \theta_2, \theta_1}.$
Any path integral over field configurations on a wedge of width $\theta_2$ can be expressed as a path integral over field configurations on a wedge of width $\theta_1,$ provided we transform both the fields \textit{and the metric} by $\psi_{f, \theta_2, \theta_1}.$
Concretely if $\eta_{\mu \nu}$ is the Minkowski metric and $D_{g}[\phi]$ is the path integral measure with respect to the metric $g$, we have
\begin{align}
	\begin{split}
	\langle \varphi_2 | W_{\theta_2} | \varphi_1\rangle
		& = \int_{\phi(\theta=0) = \varphi_1}^{\phi(\theta = \theta_2) = \varphi_2 \circ R_{-\theta_2}} D_\eta[\phi] e^{- S^{E}_{\eta}[\phi]} \\
		& = \int_{\phi(\theta=0) = \varphi_1}^{\phi(\theta = \theta_2) = \varphi_2 \circ R_{-\theta_2}} D_{\psi_{f, \theta_2, \theta_1}^* \eta}[\phi \circ \psi_{f, \theta_2, \theta_1}^{-1}] e^{- S^E_{\psi_{f, \theta_2, \theta_1}^* \eta}[\phi \circ \psi_{f, \theta_2, \theta_1}^{-1}]}
	\end{split}
\end{align}
We can simply relabel our integration variable $\phi \circ \psi_{f, \theta_2, \theta_1}^{-1} \mapsto \phi$, and we obtain
\begin{align}
	\begin{split}
		\langle \varphi_2 | W_{\theta_2} | \varphi_1\rangle
		& = \int_{\phi(\theta=0) = \varphi_1}^{\phi(\theta = \theta_1) = \varphi_2 \circ R_{-\theta_1}} D_{\psi_{f, \theta_2, \theta_1}^* \eta}[\phi] e^{- S^E_{\psi_{f, \theta_2, \theta_1}^* \eta}[\phi]}.
	\end{split}
\end{align}
So understanding the relationship between $W_{\theta_1}$ and $W_{\theta_2}$ is all about understanding how the path integral measure and the action change under a small deformation of the metric.
For infinitesimal transformations, this is what \textit{defines} the stress-energy tensor.
When we take $\theta_2 \to \theta_1,$ denoting $X^\mu = f(\theta) \left(\frac{\del}{\del \theta} \right)^\mu,$ we get
\begin{equation}
	(\psi^*_{f, \theta_2, \theta_1} \eta)_{\mu \nu} \sim \eta_{\mu \nu} + (\theta_2 - \theta_1) (\del_\mu X_\nu + \del_\nu X_\mu).
\end{equation}
Under a small change in the metric, we have
\begin{equation}
	D_{g + \delta g}[\phi] e^{- S^E_{g + \delta g}[\phi]}
		\equiv D_{g}[\phi] e^{- S^E_{g}[\phi]} \left( 1 - \frac{1}{2} \int d^{d+1} x \sqrt{|g|} T_E^{ab} \delta g_{ab}\right)
\end{equation}
From this, and our above expressions, we can easily compute
\begin{align}
	\begin{split}
		\frac{\del}{\del \theta_1}
		\langle \varphi_2 | W_{\theta_1} | \varphi_1\rangle
		 = - \int_{\phi(\theta=0) = \varphi_1}^{\phi(\theta = \theta_1) = \varphi_2 \circ R_{-\theta_1}} D_{\eta}[\phi] e^{- S^E_{\eta}[\phi]} \int d^{d+1} x T^E_{\mu \nu} \Del^{\mu} X^{\nu}.
	\end{split}
\end{align}
This is essentially some rather complicated time-ordered, smeared one-point function of the operator $T_{\mu \nu}^{E }\Del^{\mu} X^{\nu}.$
But because any time-ordered, smeared one-point function of the stress tensor is divergenceless, we may integrate by parts and obtain
\begin{align}
	\begin{split}
		\frac{\del}{\del \theta_1}
		\langle \varphi_2 | W_{\theta_1} | \varphi_1\rangle
		= - \int_{\phi(\theta=0) = \varphi_1}^{\phi(\theta = \theta_1) = \varphi_2 \circ R_{-\theta_1}} D_{\eta}[\phi] e^{- S^E_{\eta}[\phi]} 
		\left( \int_{\Sigma_{\theta_1}} T^E_{\mu \nu} X^{\mu} d\Sigma^{\nu} - \int_{\Sigma_{0}} T^E_{\mu \nu} X^{\mu} d\Sigma^{\nu} \right)
	\end{split}
\end{align}
The vector field $X^{\mu}$ was chosen to vanish on $\Sigma_0$ and be equal to $\frac{\del}{\del \theta}$ on $\Sigma_{\theta_1},$ so we have
\begin{align}
	\begin{split}
		\frac{\del}{\del \theta_1}
		\langle \varphi_2 | W_{\theta_1} | \varphi_1\rangle
		& = - \int_{\phi(\theta=0) = \varphi_1}^{\phi(\theta = \theta_1) = \varphi_2 \circ R_{-\theta_1}} D_{\eta}[\phi] e^{- S^E_{\eta}[\phi]} 
		\int_{\Sigma_{\theta_1}} T^E_{\mu \nu} \left(\frac{\del}{\del \theta_1}\right)^{\mu} d\Sigma^{\nu} \\
		& = - \langle \varphi_2 | \left( \int_{\Sigma_{\theta_1}} T^E_{\mu \nu} \left(\frac{\del}{\del \theta_1}\right)^{\mu} d\Sigma^{\nu} \right) W_{\theta_1} |\varphi_1\rangle.
	\end{split}
\end{align}
This clearly matches our desired equation \eqref{eq:appendix-desired-equation} with $K_{\theta}$ defined as in equation \eqref{eq:angle-generator} in the main text.

\bibliographystyle{JHEP}
\bibliography{bibliography}

\end{document}